\documentclass[english,conference, 10pt, onecolumn]{IEEEtran}
\usepackage[T1]{fontenc}
\usepackage[latin9]{inputenc}
\usepackage{verbatim}
\usepackage{float}
\usepackage{amsmath}
\usepackage{amssymb}
\usepackage{graphicx}

\makeatletter

\providecommand{\tabularnewline}{\\}
\floatstyle{ruled}
\newfloat{algorithm}{tbp}{loa}
\providecommand{\algorithmname}{Algorithm}
\floatname{algorithm}{\protect\algorithmname}

\IEEEoverridecommandlockouts

\newtheorem{theorem}{Theorem}
\newtheorem{lemma}{Lemma}
\newtheorem{definition}{Definition}

\usepackage{cite,algorithmic}

\makeatother

\usepackage{babel}
\begin{document}
\global\long\def\et#1{\tilde{\mathbf{e}}_{#1,h_{#1}}}
\global\long\def\gt{\tilde{\mathbf{G}}_{h_{1},h_{2}}}
\global\long\def\gh{\tilde{\mathbf{G}}_{h}}
\global\long\def\ghl{\tilde{\mathbf{G}}_{h(l)}}

\global\long\def\ft#1{\tilde{\mathbf{f}}_{#1,h_{#1}}}
\global\long\def\fth#1{\tilde{\mathbf{f}}_{#1,h}}
\global\long\def\ftl#1{\tilde{\mathbf{f}}_{#1,h(l)}}
\global\long\def\et#1{\tilde{\mathbf{e}}_{#1,h_{#1}}}
\global\long\def\eth#1{\tilde{\mathbf{e}}_{#1,h}}

\global\long\def\bE{\mathbb{E}}
\global\long\def\ez{\mathbb{E}_{\mathbf{Z}}}
\global\long\def\var{\mathbb{V}}
\global\long\def\bias{\mathbb{B}}
\global\long\def\S{\mathcal{S}}



\author{\IEEEauthorblockN{Kevin R. Moon\IEEEauthorrefmark{1}, Kumar Sricharan\IEEEauthorrefmark{2}, Kristjan Greenewald\IEEEauthorrefmark{3}, Alfred O. Hero III\IEEEauthorrefmark{4}} \IEEEauthorblockA{\IEEEauthorrefmark{1}Genetics Dept. and Applied Math Program, Yale University,  kevin.moon@yale.edu} \IEEEauthorblockA{\IEEEauthorrefmark{2}Intuit, sricharan$\textunderscore$kumar@intuit.com} 
\IEEEauthorblockA{\IEEEauthorrefmark{3}MIT-IBM Watson AI Lab, Kristjan.H.Greenewald@ibm.com}\IEEEauthorblockA{\IEEEauthorrefmark{4}EECS Dept., University of Michigan, hero@eecs.umich.edu}  \thanks{This work was partially supported by ARO MURI grant W911NF-15-1-0479, NSF grant CCF-1217880, and a NSF Graduate Research Fellowship to the first author under Grant No. F031543. This paper appeared in part in the Proceedings of the 2016 IEEE Intl. Symposium on Information Theory (ISIT)~\cite{moon2016improving}.}}

\title{Ensemble Estimation of Information Divergence}
\maketitle
\begin{abstract}
Recent work has focused on the problem of nonparametric estimation
of information divergence functionals. Many existing approaches are
restrictive in their assumptions on the density support set or require
difficult calculations at the support boundary which must be known
\emph{a priori}. The MSE convergence rate of a leave-one-out kernel
density plug-in divergence functional estimator for general bounded
density support sets is derived where knowledge of the support boundary
is not required. The theory of optimally weighted ensemble estimation
is generalized to derive a divergence estimator that achieves the
parametric rate when the densities are sufficiently smooth. The asymptotic
distribution of this estimator and some guidelines for tuning parameter
selection are provided. Based on the theory, an empirical estimator
of R\'enyi-$\alpha$ divergence is proposed that outperforms the
standard kernel density plug-in estimator, especially in high dimension.
The estimator is shown to be robust to the choice of tuning parameters.
As an illustration, we use the estimator to estimate bounds on the
Bayes error rate of a classification problem.
\end{abstract}


\section{Introduction}

Information divergences are integral functionals of two probability
distributions and have many applications in the fields of information
theory, statistics, signal processing, and machine learning. Some
applications of divergences include estimating the decay rates of
error probabilities~\cite{cover2012elements}, estimating bounds
on the Bayes error for a classification problem~\cite{avi1996bound,hashlamoun1994bound,moon2015Bayes,chernoff1952measure,berisha2014bound,moon2014nips,gliske2015intrinsic},
extending machine learning algorithms to distributional features~\cite{poczos2011estimation,oliva2013distribution,szabo2014distribution,moon2015partII},
testing the hypothesis that two sets of samples come from the same
probability distribution~\cite{moon2015partI}, clustering~\cite{dhillon2003cluster,banerjee2005clustering,lewi2006real},
feature selection and classification~\cite{bruzzone1995feature,guorong1996feature,sakate2014variable},
blind source separation~\cite{hild2001blind,mihoko2002blind}, image
segmentation~\cite{vemuri2011segment,hamza2003segmentation,liu2014segment},
and steganography~\cite{korzhik2015steganographic}. For many more
applications of divergence measures, see~\cite{basseville2013divergence}.
An important subset of information divergences is the family of $f$-divergences~\cite{csiszar1967div,ali1966div}.
This family includes the well-known Kullback-Leibler (KL) divergence~\cite{kullback1951divergence},
the R\'{e}nyi-$\alpha$ divergence~\cite{renyi1961divergence},
the Hellinger-Bhattacharyya distance~\cite{hellinger1909,bhattacharyya1946div},
the Chernoff-$\alpha$ divergence~\cite{chernoff1952measure}, the
total variation distance, and the Henze-Penrose divergence~\cite{berisha2014bound}. 

Despite the many applications of divergences, no nonparametric estimators
of these functionals exist that achieve the parametric mean squared
error (MSE) convergence rate, are simple to implement, do not require
knowledge of the boundary of the densities' support set, and apply
to a large set of divergence functionals. In this paper, we present
the first information divergence estimator that achieves all of the
above. Specifically, we consider the problem of estimating divergence
functionals when only a finite population of independent and identically
distributed (i.i.d.) samples is available from the two $d$-dimensional
distributions that are unknown, nonparametric, and smooth. Our contributions
are
\begin{enumerate}
\item We propose the first information divergence estimator, referred to
as EnDive, that is based on ensemble methods. The ensemble estimator
takes a weighted average of an ensemble of weak kernel density plug-in
estimators of divergence where the weights are chosen to improve the
MSE convergence rate. This ensemble construction makes it very easy
to implement EnDive.
\item We prove that the proposed ensemble divergence estimator achieves
the optimal parametric MSE rate of $O\left(\frac{1}{N}\right)$, where
$N$ is the sample size, when the densities are sufficiently smooth.
In particular, EnDive achieves these rates without explicitly performing
boundary correction, which is required for most other estimators.
Furthermore, we show that the convergence rates are uniform.
\item We prove that EnDive obeys a central limit theorem and thus can be
used to perform inference tasks on the divergence such as testing
that two populations have identical distributions or constructing
confidence intervals.
\end{enumerate}

\subsection{Related Work}

While several estimators of divergence functionals have been previously
defined, the convergence rates are known for only a few of them. Furthermore,
the asymptotic distributions of these estimators is unknown for nearly
all of them. For example, P\'{o}czos and Schneider~\cite{poczos2011estimation}
established weak consistency of a bias-corrected $k$-nn estimator
for R\'{e}nyi-$\alpha$ and other divergences of a similar form where
$k$ is fixed. Wang et al~\cite{wang2009divergence} provided a $k$-nn
based estimator for the KL divergence. Mutual information and divergence
estimators based on plug-in histogram schemes have been proven to
be consistent~\cite{darbellay1999MIest,silva2010partition,le2013partition,wang2005part}.
Hero et al~\cite{hero2002applications} provided an estimator for
R\'{e}nyi-$\alpha$ divergence but assumed that one of the densities
was known. However none of these works study the convergence rates
nor the asymptotic distribution of their estimators.

There has been recent interest in deriving convergence rates for divergence
estimators~\cite{moon2014isit,nguyen2010div,krishnamurthy2014divergence,singh2014renyi,singh2014exponential,kandasamy2015nonparametric}.
The rates are typically derived in terms of a smoothness condition
on the densities, such as the H\"{o}lder condition~\cite{hardle1990applied}:

\begin{definition}[H\"{o}lder Class] \label{def:holder}\emph{Let
$\mathcal{X}\subset\mathbb{R}^{d}$ be a compact space. For $r=(r_{1},\dots,r_{d}),$
$r_{i}\in\mathbb{N},$ define $|r|=\sum_{i=1}^{d}r_{i}$ and $D^{r}=\frac{\partial^{|r|}}{\partial x_{1}^{r_{1}}\dots\partial x_{d}^{r_{d}}}$.
The H\"{o}lder class $\Sigma(s,K_{H})$ of functions on $L_{2}(\mathcal{X})$
consists of the functions $f$ that satisfy 
\[
\left|D^{r}f(x)-D^{r}f(y)\right|\leq K_{H}\left\Vert x-y\right\Vert ^{s-|r|},
\]
for all $x,\,y\in\mathcal{X}$ and for all $r$ s.t. $|r|\leq\left\lfloor s\right\rfloor $.
}\end{definition}

From Definition~\ref{def:holder}, it is clear that if a function
$f$ belongs to $\Sigma(s,K_{H})$, then $f$ is continuously differentiable
up to order $\left\lfloor s\right\rfloor $. In this work, we show
that EnDive achieves the parametric MSE convergence rate of $O(1/N)$
when $s\geq d$ and $s>\frac{d}{2}$, depending on the specific form
of the divergence function.

Nguyen et al.~\cite{nguyen2010div} proposed a method for estimating
$f$-divergences by estimating the likelihood ratio of the two densities
by solving a convex optimization problem and then plugging it into
the divergence formulas. For this method the authors prove that the
minimax MSE convergence rate is parametric when the likelihood ratio
is in the bounded H\"{o}lder class $\Sigma(s,K_{H})$ with $s\geq d/2$.
However, this estimator is restricted to true $f$-divergences and
may not apply to the broader class of divergence functionals that
we consider here (as an example, the $L_{2}^{2}$ divergence is not
an $f$-divergence). Additionally, solving the convex problem of~\cite{nguyen2010div}
has similar computational complexity to that of training a support
vector machine (SVM) (between $O(N^{2})$ and $O(N^{3})$), which
can be demanding when $N$ is large. In contrast, EnDive depends only
on simple density plug-in estimates and the solution of an offline
convex optimization problem. Thus the most computationally demanding
step in our approach is the calculation of the density estimates,
which has complexity no greater than $O(N^{2})$.

Singh and P\'{o}czos~\cite{singh2014renyi,singh2014exponential}
provided an estimator for R\'{e}nyi-$\alpha$ divergences as well
as general density functionals that uses a ``mirror image'' kernel
density estimator. They prove a convergence rate of $O\left(\frac{1}{N}\right)$
when $s\geq d$ for each of the densities. However this method requires
several computations at each boundary of the support of the densities
which is difficult to implement as $d$ gets large. Also, this method
requires knowledge of the support of the densities which is unknown
in most practical settings. In contrast, while our assumptions require
the density support sets to be bounded, knowledge of the support is
not required for implementation.

The ``linear'' and ``quadratic'' estimators presented by Krishnamurthy
et al~\cite{krishnamurthy2014divergence} estimate divergence functionals
that include the form $\int f_{1}^{\alpha}(x)f_{2}^{\beta}(x)d\mu(x)$
for given $\alpha$ and $\beta$ where $f_{1}$ and $f_{2}$ are probability
densities. These estimators achieve the parametric rate when $s\geq d/2$
and $s\geq d/4$ for the linear and quadratic estimators, respectively.
However, the latter estimator is computationally infeasible for most
functionals and the former requires numerical integration for some
divergence functionals, which can be computationally difficult. Additionally,
while a suitable $\alpha$-$\beta$ indexed sequence of divergence
functionals of this form can be made to converge to the KL divergence,
this does not guarantee convergence of the corresponding sequence
of divergence estimators in~\cite{krishnamurthy2014divergence},
whereas our estimator can be used to directly estimate the KL divergence.
Other important $f$-divergence functionals are also excluded from
this form including some that bound the Bayes error~\cite{moon2015Bayes,berisha2014bound,avi1996bound}.
In contrast, our method applies to a large class of divergence functionals
and avoids numerical integration.

Finally, Kandasamy et al.~\cite{kandasamy2015nonparametric} propose
influence function based estimators of distributional functionals
including divergences that achieve the parametric rate when $s\geq d/2$.
While this method can be applied to general functionals, the estimator
requires numerical integration for some functionals. Additionally,
the estimators in both Kandasamy et al~\cite{kandasamy2015nonparametric}
and Krishnamurthy et al~\cite{krishnamurthy2014divergence} require
an optimal kernel density estimator. This is difficult to construct
when the density support is bounded as it requires knowledge of the
densities' support boundary and difficult computations at the boundary,
whereas our method does not require knowledge of the support boundary.

Asymptotic normality has been established for certain appropriately
normalized divergences between a specific density estimator and the
true density~\cite{berlinet1995l1,berlinet1997asymptotic,bickel1973some}.
This differs from our setting where we assume that both densities
are unknown. The asymptotic distributions of the estimators in~\cite{krishnamurthy2014divergence,singh2014exponential,singh2014renyi,nguyen2010div}
are currently unknown. Thus it is difficult to use these estimators
for hypothesis testing, which is crucial in many scientific applications.
Kandasamy et al~\cite{kandasamy2015nonparametric} prove a central
limit theorem for their data-splitting estimator but do not prove
similar results for their leave-one-out estimator. We establish a
central limit theorem for EnDive, which greatly enhances its applicability
in scientific settings.

Our ensemble divergence estimator reduces to an ensemble entropy estimator
as a special case when data from only one distribution is considered
and the other density is set to a uniform measure. The resultant entropy
estimator differs from the ensemble entropy estimator proposed by
Sricharan et al.~\cite{sricharan2013ensemble} in several important
ways. First, the densities' support set must be known for the estimator
in~\cite{sricharan2013ensemble} to perform explicit boundary correction.
In contrast, the EnDive estimator does not require any boundary correction.
To show this requires a significantly different approach in proving
the bias and variance rates of the EnDive estimator. Furthermore,
the EnDive results apply under more general assumptions on the densities
and the kernel used in the weak estimators. Finally, the central limit
theorem applies to the EnDive estimator which is currently unknown
for the estimator in~\cite{sricharan2013ensemble}.

\subsection{Organization and Notation}

The paper is organized as follows. We first derive MSE convergence
rates in Section~\ref{sec:base_est} for a weak divergence estimator,
which is a kernel density plug-in divergence estimator. We then generalize
the theory of optimally weighted ensemble entropy estimation developed
in~\cite{sricharan2013ensemble} to obtain the ensemble divergence
estimator EnDive from an ensemble of weak estimators in Section~\ref{sec:weighted}.
A central limit theorem and uniform convergence rate for the ensemble
estimator are also presented in Section~\ref{sec:weighted}. In Section~\ref{sec:experiments},
we provide guidelines for selecting the tuning parameters based on
experiments and the theory derived in the previous sections. We then
perform experiments in Section~\ref{sec:experiments} that validate
the theory and establish the robustness of the proposed estimators
to the tuning parameters. 

Bold face type is used for random variables and random vectors. The
conditional expectation given a random variable $\mathbf{Z}$ is denoted
$\mathbb{E}_{\mathbf{Z}}$. The variance of a random variable is denoted
$\var$ and the bias of an estimator is denoted $\bias$.

\section{The Divergence Functional Weak Estimator}

\label{sec:base_est}This paper focuses on estimating functionals
of the form
\begin{equation}
G\left(f_{1},f_{2}\right)=\int g\left(f_{1}(x),f_{2}(x)\right)f_{2}(x)dx,\label{eq:fdiv}
\end{equation}
where $g(x,y)$ is a smooth functional, and $f_{1}$ and $f_{2}$
are smooth $d$-dimensional probability densities. If $g\left(f_{1}(x),f_{2}(x)\right)=g\left(\frac{f_{1}(x)}{f_{2}(x)}\right),$
$g$ is convex, and $g(1)=0$, then $G\left(f_{1},f_{2}\right)$ defines
the family of $f$-divergences. Some common divergences that belong
to this family include the KL divergence ($g(t)=-\ln t$) and the
total variation distance ($g(t)=|t-1|$). In this work, we consider
a broader class of functionals than the $f$-divergences since $g$
is allowed to be very general.

To estimate $G\left(f_{1},f_{2}\right)$, we will first define a weak
plug-in estimator based on kernel density estimators (KDEs); that
is, a simple estimator that converges slowly to the true value $G\left(f_{1},f_{2}\right)$
in terms of MSE. We will then derive the bias and variance expressions
for this weak estimator as a function of sample size and bandwidth.
We then use the resulting bias and variance expressions to derive
an ensemble estimator that takes a weighted average of weak estimators
with different bandwidths and achieves superior MSE performance.

\subsection{The Kernel Density Plug-in Estimator}

We use a kernel density plug-in estimator of the divergence functional
in (\ref{eq:fdiv}) as the weak estimator. Assume that $N_{1}$ i.i.d.
realizations $\left\{ \mathbf{Y}_{1},\dots,\mathbf{Y}_{N_{1}}\right\} $
are available from $f_{1}$ and $N_{2}$ i.i.d. realizations $\left\{ \mathbf{X}_{1},\dots,\mathbf{X}_{N_{2}}\right\} $
are available from $f_{2}$. Let $h_{i}>0$ be the kernel bandwidth
for the density estimator of $f_{i}$. For simplicity of presentation,
assume that $N_{1}=N_{2}=N$ and $h_{1}=h_{2}=h$. The results for
the more general case of differing sample sizes and bandwidths are
given in Appendix~\ref{sec:general}. Let $K(\cdot)$ be a kernel
function with $\int K(x)dx=1$ and $||K||_{\infty}<\infty$ where
$\|K\|_{\infty}$ is the $\ell_{\infty}$ norm of the kernel $K$.
The KDEs for $f_{1}$ and $f_{2}$ are, respectively, 
\begin{eqnarray*}
\fth 1(\mathbf{X}_{j}) & = & \frac{1}{Nh^{d}}\sum_{i=1}^{N}K\left(\frac{\mathbf{X}_{j}-\mathbf{Y}_{i}}{h}\right),\\
\fth 2(\mathbf{X}_{j}) & = & \frac{1}{Mh^{d}}\sum_{\substack{i=1\\
i\neq j
}
}^{N}K\left(\frac{\mathbf{X}_{j}-\mathbf{X}_{i}}{h}\right),
\end{eqnarray*}
where $M=N-1$. $G\left(f_{1},f_{2}\right)$ is then approximated
as 
\begin{equation}
\gh=\frac{1}{N}\sum_{i=1}^{N}g\left(\fth 1\left(\mathbf{X}_{i}\right),\fth 2\left(\mathbf{X}_{i}\right)\right).\label{eq:estimator}
\end{equation}

\subsection{Convergence Rates}

For many estimators, MSE convergence rates are typically provided
in the form of upper (or sometimes lower) bounds on the bias and the
variance. Therefore, only the slowest converging terms (as a function
of sample size $N$) are presented in these cases. However, to apply
our generalized ensemble theory to obtain estimators that guarantee
the parametric MSE rate, we require explicit expressions for the bias
of the weak estimators in terms of the sample size $N$ and the kernel
bandwidth $h$. Thus an upper bound is insufficient for our work.
Furthermore, to guarantee the parametric rate, we require explicit
expressions of all bias terms that converge to zero slower than $O\left(1/\sqrt{N}\right)$.

To obtain such expressions, we require multiple assumptions on the
densities $f_{1}$ and $f_{2}$, the functional $g$, and the kernel
$K$. Similar to~\cite{sricharan2013ensemble,moon2014isit,moon2014nips},
the principal assumptions we make are that: 1) $f_{1}$, $f_{2},$
and $g$ are smooth; 2) $f_{1}$ and $f_{2}$ have common bounded
support sets $\mathcal{S}$; 3) $f_{1}$ and $f_{2}$ are strictly
lower bounded on $\S$. We also assume 4) that the densities' support
set is smooth with respect to the kernel $K(u)$. Our full assumptions
are:
\begin{itemize}
\item $(\mathcal{A}.0)$: Assume that the kernel $K$ is symmetric, is a
product kernel, and has bounded support in each dimension. 
\item $(\mathcal{A}.1)$: Assume there exist constants $\epsilon_{0},\epsilon_{\infty}$
such that $0<\epsilon_{0}\leq f_{i}(x)\leq\epsilon_{\infty}<\infty,\,\forall x\in S.$ 
\item $(\mathcal{A}.2)$: Assume that the densities $f_{i}\in\Sigma(s,K_{H})$
in the interior of $\mathcal{S}$ with $s\geq2$.
\item $(\mathcal{A}.3)$: Assume that $g$ has an infinite number of mixed
derivatives.
\item $(\mathcal{A}.4$): Assume that $\left|\frac{\partial^{k+l}g(x,y)}{\partial x^{k}\partial y^{l}}\right|$,
$k,l=0,1,\ldots$ are strictly upper bounded for $\epsilon_{0}\leq x,y\leq\epsilon_{\infty}$. 
\item $(\mathcal{A}.5)$: Assume the following boundary smoothness condition:
Let $p_{x}(u):\mathbb{R}^{d}\rightarrow\mathbb{R}$ be a polynomial
in $u$ of order $q\leq r=\left\lfloor s\right\rfloor $ whose coefficients
are a function of $x$ and are $r-q$ times differentiable. Then assume
that 
\[
\int_{x\in\mathcal{S}}\left(\int_{u:K(u)>0,\,x+uh\notin\mathcal{S}}K(u)p_{x}(u)du\right)^{t}dx=v_{t}(h),
\]
where $v_{t}(h)$ admits the expansion 
\[
v_{t}(h)=\sum_{i=1}^{r-q}e_{i,q,t}h^{i}+o\left(h^{r-q}\right),
\]
for some constants $e_{i,q,t}$.
\end{itemize}
We focus on finite support kernels for simplicity in the proofs although
it is likely that our results extend to some infinitely supported
kernels as well. We assume relatively strong conditions on the smoothness
of $g$ in $\mathcal{A}.3$ to enable us to obtain an estimator that
achieves good convergence rates without knowledge of the boundary
of the support set. While this smoothness condition may seem restrictive,
in practice nearly all divergence and entropy functionals of interest
satisfy this condition. Functionals of interest that do not satisfy
this assumption (e.g. the total variation distance) typically have
at least one point that is not differentiable which violates the assumptions
of all competing estimators~\cite{sricharan2013ensemble,moon2014isit,kandasamy2015nonparametric,krishnamurthy2014divergence,singh2014exponential,singh2014renyi}.
We also note that to obtain simply an upper bound on the bias for
the plug-in estimator, much less restrictive assumptions on the functional
$g$ are sufficient. 

Assumption $\mathcal{A}.5$ requires the boundary of the density support
set to be smooth with respect to the kernel $K(u)$ in the sense that
the expectation of the area outside of $\mathcal{S}$ with respect
to any random variable $u$ with smooth distribution is a smooth function
of the bandwidth $h$. It is not necessary for the boundary of $\mathcal{S}$
to have smooth contours with no edges or corners as this assumption
is satisfied by the following case:

\begin{theorem}\label{thm:boundary}\emph{Assumption $\mathcal{A}.5$
is satisfied when $\mathcal{S}=[-1,1]^{d}$ and when $K$ is the uniform
rectangular kernel; that is $K(x)=1$ for all $x:\,||x||_{1}\leq1/2$.}

\end{theorem}

The proof is given in Appendix~\ref{sec:boundaryProof}. Given the
simple nature of this density support set and kernel, it is likely
that other kernels and supports will satisfy $\mathcal{A}.5$ as well.
This is left for future work. Note that assumption $\mathcal{A}.0$
is trivially satisfied by the uniform rectangular kernel as well.

Densities for which assumptions $\mathcal{A}.1-\mathcal{A}.2$ hold
include the truncated Gaussian distribution and the Beta distribution
on the unit cube. Functions for which assumptions $\mathcal{A}.3-\mathcal{A}.4$
hold include $g(x,y)=-\ln\left(\frac{x}{y}\right)$ and $g(x,y)=\left(\frac{x}{y}\right)^{\alpha}.$
The following theorem on the bias follows under assumptions $\mathcal{A}.0-\mathcal{A}.5$:

\begin{theorem}

\label{thm:bias}\emph{For general $g$, the bias of the plug-in estimator
$\gh$ is given by 
\begin{equation}
\bias\left[\gh\right]=\sum_{j=1}^{\left\lfloor s\right\rfloor }c_{10,j}h^{j}+c_{11}\frac{1}{Nh^{d}}+O\left(h^{s}+\frac{1}{Nh^{d}}\right).\label{eq:bias1}
\end{equation}
}

\end{theorem}

To apply our generalized ensemble theory to the KDE plug-in estimator
\emph{$\gh$}, we require only an upper bound on its variance. The
following variance result requires much less strict assumptions than
the bias results in Theorem~\ref{thm:bias}:

\begin{theorem}\label{thm:variance}\emph{Assume that the functional
$g$ in~}(\ref{eq:fdiv})\emph{ is Lipschitz continuous in both of
its arguments with Lipschitz constant $C_{g}$. Then the variance
of the plug-in estimator $\gh$ is bounded by 
\[
\var\left[\gh\right]\leq C_{g}^{2}||K||_{\infty}^{2}\frac{11}{N}.
\]
}

\end{theorem}

From Theorems~\ref{thm:bias} and~\ref{thm:variance}, it is clear
that we require $h\rightarrow0$ and $Nh^{d}\rightarrow\infty$ for
$\gh$ to be unbiased while the variance of the plug-in estimator
depends primarily on the number of samples. Note that the constants
depend on the densities $f_{1}$ and $f_{2}$ and their derivatives
which are often unknown. 

\subsection{Optimal MSE Rate}

\label{subsec:optimal}From Theorem~\ref{thm:bias}, the dominating
terms in the bias are $\Theta\left(h\right)$ and $\Theta\left(\frac{1}{Nh^{d}}\right)$.
If no attempt is made to correct the bias, the optimal choice of $h$
in terms of minimizing the MSE is 
\[
h^{*}=\Theta\left(N^{\frac{-1}{d+1}}\right).
\]
 This results in a dominant bias term of order $\Theta\left(N^{\frac{-1}{d+1}}\right)$.
Note that this differs from the standard result for the optimal KDE
bandwidth for minimum MSE density estimation which is $\Theta\left(N^{-1/\left(d+4\right)}\right)$
for a symmetric uniform kernel~\cite{hansen2009lecture}.

\begin{figure}
\centering

\includegraphics[width=0.5\textwidth]{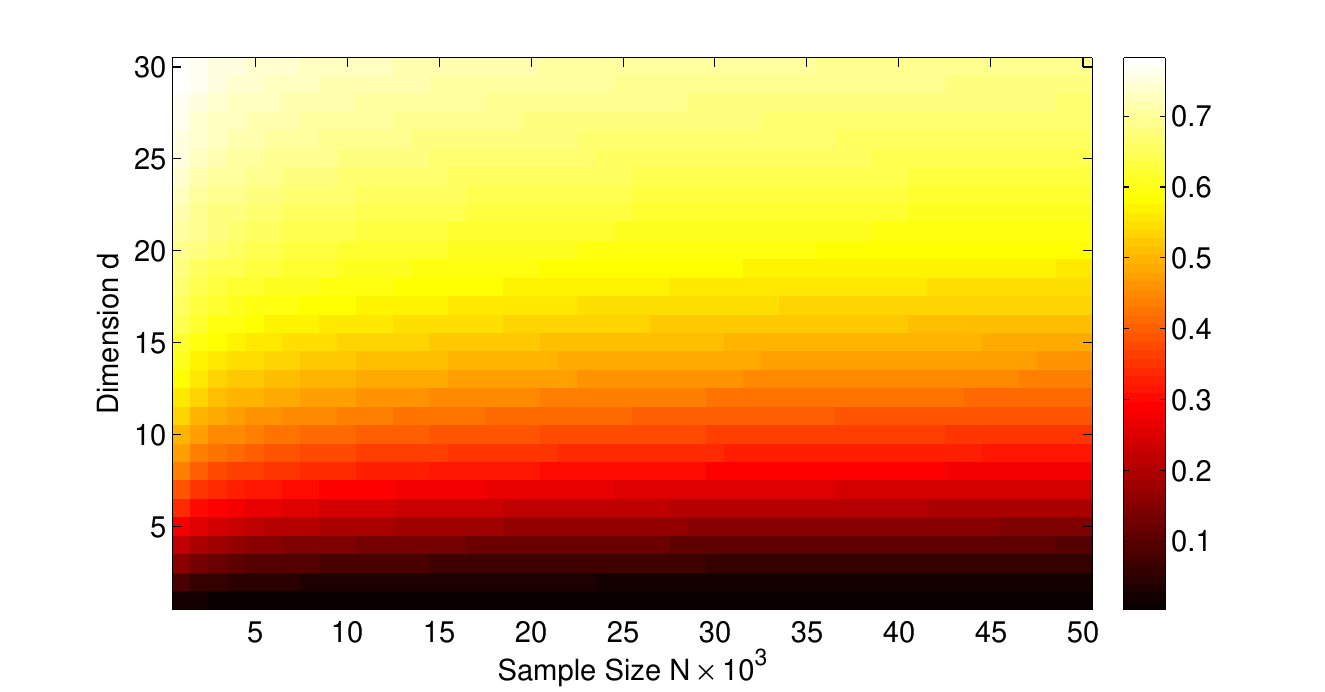}

\caption{Heat map of predicted bias of divergence funtional plug-in estimator
based on Theorem~\ref{thm:bias} as a function of dimension and sample
size when $h=N^{\frac{-1}{d+1}}$. Note the phase transition in the
bias as dimension $d$ increases for fixed sample size $N$: bias
remains small only for relatively small values of $d.$ The proposed
weighted ensemble estimator removes this phase transition when the
densities are sufficiently smooth. \label{fig:heatmap}}
\end{figure}

Figure~\ref{fig:heatmap} shows a heatmap showing the leading bias
term $O\left(h\right)$ as a function of $d$ and $N$ when $h=N^{\frac{-1}{d+1}}$.
The heatmap indicates that the bias of the plug-in estimator in (\ref{eq:estimator})
is small only for relatively small values of $d$. In the next section,
we propose an ensemble estimator that achieves a superior convergence
rate regardless of dimension $d$ as long as the densities are sufficiently
smooth.

\subsection{Proof Sketches of Theorems~\ref{thm:bias} and~\ref{thm:variance}}

\label{sec:Proofs}To prove the expressions for the bias, the bias
is first decomposed into two parts by adding and subtracting $g\left(\ez\fth 1(\mathbf{Z}),\ez\fth 2(\mathbf{Z})\right)$
within the expectation creating a ``bias'' term and a ``variance''
term. Applying a Taylor series expansion on the bias and variance
terms results in expressions that depend on powers of $\bias_{\mathbf{Z}}\left[\fth i(\mathbf{Z})\right]:=\ez\fth i(\mathbf{Z})-f_{i}(\mathbf{Z})$
and $\eth i(\mathbf{Z}):=\fth i(\mathbf{Z})-\ez\fth i(\mathbf{Z})$,
respectively. Within the interior of the support, moment bounds can
be derived from properties of the KDEs and a Taylor series expansion
of the densities. Near the boundary of the support, the smoothness
assumption on the boundary $\mathcal{A}.5$ is required to obtain
an expression of the bias in terms of the KDE bandwidth $h$ and the
sample size $N$. The full proof of Thm.~\ref{thm:bias} is given
in Appendix~\ref{sec:BiasProof}.

The proof of the variance result takes a different approach. The proof
uses the Efron-Stein inequality~\cite{efron1981jackknife} which
bounds the variance by analyzing the expected squared difference between
the plug-in estimator when one sample is allowed to differ. This approach
provides a bound on the variance under much less strict assumptions
on the densities and the functional $g$ than is required for Theorem~\ref{thm:bias}.
The full proof of Thm.~\ref{thm:variance} is given in Appendix~\ref{sec:VarProof}.

\section{Weighted Ensemble Estimation}

\label{sec:weighted}Theorem.~\ref{thm:bias} shows that when the
dimension of the data is not small, the bias of the MSE-optimal plug-in
estimator $\gh$ decreases very slowly as a function of sample size,
resulting in large MSE. However, by applying the theory of optimally
weighted ensemble estimation, we can modify the minimum MSE estimator
by taking a weighted sum of an ensemble of estimators where the weights
are chosen to significantly reduce the bias.

We form an ensemble of estimators by choosing different values of
$h$. Choose $\mathcal{L}=\left\{ l_{1},\dots,l_{L}\right\} $ to
be real positive numbers that index $h(l_{i})$. Thus the parameter
$l$ indexes over different neighborhood sizes for the KDEs. Define
$w:=\left\{ w\left(l_{1}\right),\dots,w\left(l_{L}\right)\right\} $
and $\tilde{\mathbf{G}}_{w}:=\sum_{l\in\mathcal{L}}w(l)\tilde{\mathbf{G}}_{h(l)}.$
The key to reducing the MSE is to choose the weight vector $w$ to
reduce the lower order terms in the bias without substantially increasing
the variance. 

\subsection{Finding the Optimal Weight}

The theory of optimally weighted ensemble estimation is a general
theory originally presented by Sricharan et al~\cite{sricharan2013ensemble}
that can be applied to estimation problems as long as the bias and
variance of the estimator can be expressed in a specific way. We now
generalize this theory so that it can be applied to a wider variety
of estimation problems. Let $\mathcal{L}=\left\{ l_{1},\dots,l_{L}\right\} $
be a set of index values and let $N$ be the number of samples available.
For an indexed ensemble of estimators $\left\{ \hat{\mathbf{E}}_{l}\right\} _{l\in\mathcal{L}}$
of a parameter $E$, the weighted ensemble estimator with weights
$w=\left\{ w\left(l_{1}\right),\dots,w\left(l_{L}\right)\right\} $
satisfying $\sum_{l\in\mathcal{L}}w(l)=1$ is defined as 
\[
\hat{\mathbf{E}}_{w}=\sum_{l\in\mathcal{L}}w\left(l\right)\hat{\mathbf{E}}_{l}.
\]
 $\hat{\mathbf{E}}_{w}$ is asyptotically unbiased if the estimators
$\left\{ \hat{\mathbf{E}}_{l}\right\} _{l\in\mathcal{L}}$ are asymptotically
unbiased. Consider the following conditions on $\left\{ \hat{\mathbf{E}}_{l}\right\} _{l\in\mathcal{L}}$: 
\begin{itemize}
\item $\mathcal{C}.1$ The bias is expressible as 
\[
\bias\left[\hat{\mathbf{E}}_{l}\right]=\sum_{i\in J}c_{i}\psi_{i}(l)\phi_{i,d}(N)+O\left(\frac{1}{\sqrt{N}}\right),
\]
where $c_{i}$ are constants depending on the underlying density and
are independent of $N$ and $l$, $J=\left\{ i_{1},\dots,i_{I}\right\} $
is a finite index set with $I<L$, and $\psi_{i}(l)$ are basis functions
depending only on the parameter $l$ and not on the sample size $N$. 
\item $\mathcal{C}.2$ The variance is expressible as 
\[
\var\left[\hat{\mathbf{E}}_{l}\right]=c_{v}\left(\frac{1}{N}\right)+o\left(\frac{1}{N}\right).
\]
\end{itemize}
\begin{theorem}\label{thm:ensemble}\emph{Assume conditions $\mathcal{C}.1$
and $\mathcal{C}.2$ hold for an ensemble of estimators $\left\{ \hat{\mathbf{E}}_{l}\right\} _{l\in\mathcal{L}}$.
Then there exists a weight vector $w_{0}$ such that the MSE of the
weighted ensemble estimator attains the parametric rate of convergence:
\[
\mathbb{E}\left[\left(\hat{\mathbf{E}}_{w_{0}}-E\right)^{2}\right]=O\left(\frac{1}{N}\right).
\]
The weight vector $w_{0}$ is the solution to the following convex
optimization problem:
\begin{equation}
\begin{array}{rl}
\min_{w} & ||w||_{2}\\
subject\,to & \sum_{l\in\mathcal{L}}w(l)=1,\\
 & \gamma_{w}(i)=\sum_{l\in\mathcal{L}}w(l)\psi_{i}(l)=0,\,i\in J.
\end{array}\label{eq:optimize}
\end{equation}
}

\end{theorem}
\begin{IEEEproof}
From condition $\mathcal{C}.1$, the bias of the weighted estimator
is 
\[
\bias\left[\hat{\mathbf{E}}_{w}\right]=\sum_{i\in J}c_{i}\gamma_{w}(i)\phi_{i,d}(N)+O\left(\frac{\sqrt{L}||w||_{2}}{\sqrt{N}}\right).
\]
The variance of the weighted estimator is bounded as 
\begin{equation}
\var\left[\hat{\mathbf{E}}_{w}\right]\leq\frac{L||w||_{2}^{2}}{N}.\label{eq:ens_var}
\end{equation}
The optimization problem in (\ref{eq:optimize}) zeroes out the lower-order
bias terms and limits the $\ell_{2}$ norm of the weight vector $w$
to limit the variance contribution. This results in an MSE rate of
$O(1/N)$ when the dimension $d$ is fixed and when $L$ is fixed
independently of the sample size $N$. Furthermore, a solution to
(\ref{eq:optimize}) is guaranteed to exist as long as $L>I$ and
the vectors $a_{i}=\left[\psi_{i}(l_{1}),\dots,\psi_{i}(l_{L})\right]$
are linearly independent. This completes our sketch of the proof of
Thm.~\ref{thm:ensemble}.
\end{IEEEproof}

\subsection{The EnDive Estimator}

\label{sub:odin}To achieve the parametric rate $O\left(1/N\right)$
in MSE convergence it is not necessary that $\gamma_{w}(i)=0,\,i\in J$.
Solving the following convex optimization problem in place of the
optimization problem in Theorem~\ref{thm:ensemble} retains the $O(1/N)$
rate:
\begin{equation}
\begin{array}{rl}
\min_{w} & \epsilon\\
subject\,to & \sum_{l\in\mathcal{L}}w(l)=1,\\
 & \left|\gamma_{w}(i)N^{\frac{1}{2}}\phi_{i,d}(N)\right|\leq\epsilon,\,\,i\in J,\\
 & \left\Vert w\right\Vert _{2}^{2}\leq\eta,
\end{array}\label{eq:relaxed}
\end{equation}
where the parameter $\eta$ is chosen to achieve a trade-off between
bias and variance. Instead of forcing $\gamma_{w}(i)=0$, the relaxed
optimization problem uses the weights to decrease the bias terms at
the rate of $O\left(1/\sqrt{N}\right)$ yielding an MSE of $O(1/N).$

We now construct a divergence ensemble estimator from an ensemble
of plug-in KDE divergence estimators. Consider first the bias result
in (\ref{eq:bias1}) where $g$ is general and assume that $s\geq d$.
In this case, the bias contains a $O\left(\frac{1}{h^{d}N}\right)$
term. To guarantee the parametric MSE rate, any remaining lower-order
bias terms in the ensemble estimator must be no slower than $O\left(1/\sqrt{N}\right)$.
Let $h(l)=lN^{-1/(2d)}$ where $l\in\mathcal{L}$. Then $O\left(\frac{1}{h(l)^{d}N}\right)=O\left(\frac{1}{l^{d}\sqrt{N}}\right)$.
We therefore obtain an ensemble of plug-in estimators $\left\{ \ghl\right\} _{l\in\mathcal{L}}$
and a weighted ensemble estimator $\tilde{\mathbf{G}}_{w}=\sum_{l\in\mathcal{L}}w(l)\tilde{\mathbf{G}}_{h(l)}$.
The bias of each estimator in the ensemble satisfies condition $\mathcal{C}.1$
with $\psi_{i}(l)=l^{i}$ and $\phi_{i,d}(N)=N^{-i/(2d)}$ for $i=1,\dots,d$.
To obtain a uniform bound on the bias with respect to $w$ and $\mathcal{L}$,
we also include the function $\psi_{d+1}(l)=l^{-d}$ with corresponding
$\phi_{d+1,d}(N)=N^{-1/2}$. The variance also satisfies condition
$\mathcal{C}.2$. The optimal weight $w_{0}$ is found by using (\ref{eq:relaxed})
to obtain an optimally weighted plug-in divergence functional estimator
$\tilde{\mathbf{G}}_{w_{0}}$ with an MSE convergence rate of $O\left(\frac{1}{N}\right)$
as long as $s\geq d$ and $L\geq d$. Otherwise, if $s<d$ we can
only guarantee the MSE rate up to $O\left(\frac{1}{N^{s/d}}\right)$.
We refer to this estimator as the \textbf{En}semble \textbf{Dive}rgence
(EnDive) estimator and denote it as $\tilde{\mathbf{G}}_{\text{EnDive}}$.

We note that for some functionals $g$ (including the KL divergence
and the Renyi-$\alpha$ divergence integral) we can modify the EnDive
estimator to obtain the parametric rate under the less strict assumption
that $s>d/2$. For details on this approach, see Appendix~\ref{sec:Mod}.

\subsection{Central Limit Theorem}

The following theorem shows that an appropriately normalized ensemble
estimator $\tilde{\mathbf{G}}_{w}$ converges in distribution to a
normal random variable under rather general conditions. Thus the same
result applies to the EnDive estimator $\tilde{\mathbf{G}}_{\text{EnDive}}$.
This enables us to perform hypothesis testing on the divergence functional.
The proof is based on the Efron-Stein inequality and an application
of Slutsky's Theorem (Appendix~\ref{sec:cltProof}). 

\begin{theorem}\label{thm:clt}\emph{Assume that the functional $g$
is Lipschitz in both arguments with Lipschitz constant $C_{g}$. Further
assume that $h(l)=o(1)$, $N\rightarrow\infty$, and $Nh(l)^{d}\rightarrow\infty$
for each $l\in\mathcal{L}$. Then for fixed $\mathcal{L}$, the asymptotic
distribution of the weighted ensemble estimator $\tilde{\mathbf{G}}_{w}$
is 
\[
Pr\left(\left(\tilde{\mathbf{G}}_{w}-\bE\left[\tilde{\mathbf{G}}_{w}\right]\right)/\sqrt{\var\left[\tilde{\mathbf{G}}_{w}\right]}\leq t\right)\rightarrow Pr(\mathbf{S}\leq t),
\]
where $\mathbf{S}$ is a standard normal random variable.}\end{theorem}

\subsection{Uniform Convergence Rates}

Here we show that the optimally weighted ensemble estimators achieve
the parametric MSE convergence rate uniformly. Denote the subset of
$\Sigma\left(s,K_{H}\right)$ with densities bounded between $\epsilon_{0}$
and $\epsilon_{\infty}$ as $\Sigma\left(s,K_{H},\epsilon_{0},\epsilon_{\infty}\right)$.

\begin{theorem}\emph{\label{thm:uniform}Let $\tilde{\mathbf{G}}_{\text{EnDive}}$
be the EnDive estimator of the functional 
\[
G(p,q)=\int g\left(p(x),q(x)\right)q(x)dx,
\]
where $p$ and $q$ are $d$-dimensional probability densities. Additionally,
let }$r=d$ and assume that $s>r$.\emph{ Then 
\[
\sup_{p,q\in\Sigma\left(s,K_{H},\epsilon_{0},\epsilon_{\infty}\right)}\bE\left[\left(\tilde{\mathbf{G}}_{w_{0}}-G(p,q)\right)^{2}\right]\leq\frac{C}{N},
\]
where $C$ is a constant.}\end{theorem}

The proof decomposes the MSE into the variance plus the square of
the bias. The variance is bounded easily by using Theorem~\ref{thm:variance}.
To bound the bias, we show that the constants in the bias terms are
continuous with respect to the densities $p$ and $q$ under an appropriate
norm. We then show that $\Sigma\left(s,K_{H},\epsilon_{0},\epsilon_{\infty}\right)$
is compact with respect to this norm and then apply an extreme value
theorem. Details are given in Appendix~\ref{sec:UniformProof}.

\section{Experimental Results}

\label{sec:experiments}In this section, we discuss the choice of
tuning parameters and validate the EnDive estimator's convergence
rates and the central limit theorem. We then use the EnDive estimator
to estimate bounds on the Bayes error for a single-cell bone marrow
data classification problem.

\subsection{Tuning Parameter Selection}

\label{sec:tuning}The optimization problem in (\ref{eq:relaxed})
has parameters $\eta$, $L$, and $\mathcal{L}$. The parameter $\eta$
provides an upper bound on the norm of the weight vector, which gives
an upper bound on the constant in the variance of the ensemble estimator.
If all the constants in (\ref{eq:bias1}) and an exact expression
for the variance of the ensemble estimator were known, then $\eta$
could be chosen to minimize the MSE. Since the constants are unknown,
by applying (\ref{eq:relaxed}), the resulting MSE of the ensemble
estimator is $O\left(\epsilon^{2}/N\right)+O\left(L\eta^{2}/N\right),$
where each term in the sum comes from the bias and variance, respectively.
Since there is a tradeoff between $\eta$ and $\epsilon$, in principle
setting $\eta=\epsilon/\sqrt{L}$ would minimize these terms asymptotically.
In practice, we find that for finite sample sizes, the variance of
the ensemble estimator is less than the upper bound of $L\eta^{2}/N$
and setting $\eta=\epsilon/\sqrt{L}$ is therefore overly restrictive.
Setting $\eta=\epsilon$ instead works well in practice.

\begin{figure}
\centering

\includegraphics[width=0.5\textwidth]{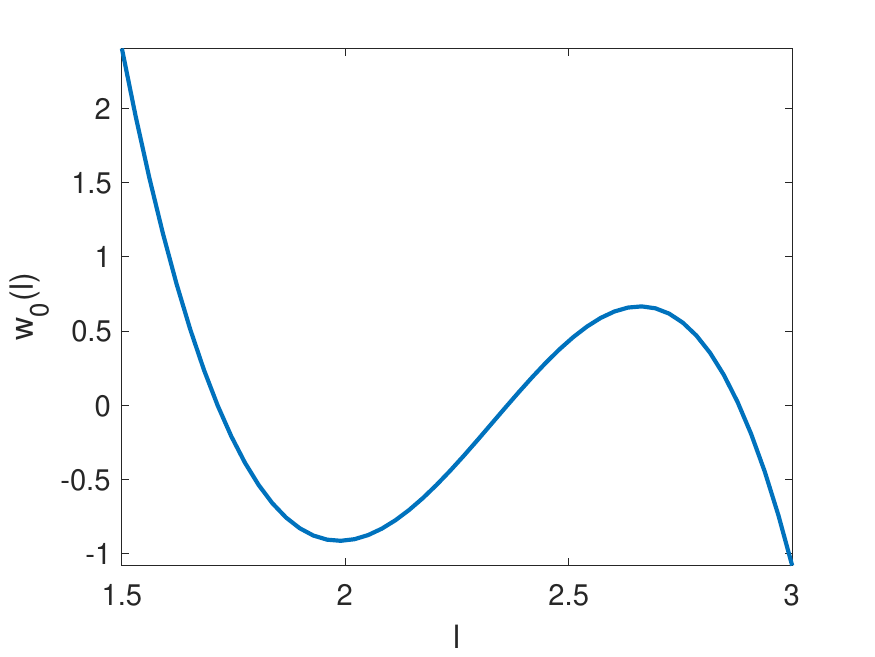}

\caption{\label{fig:weights}The optimal weights from (\ref{eq:relaxed}) when
$d=4$, $N=3100$, $L=50$, and $l$ is uniformly spaced between 1.5
and 3. The lowest values of $l$ are given the highest weight. Thus
the minimum value of bandwidth parameters $\mathcal{L}$ should be
sufficiently large to render an adequate estimate.}
\end{figure}

For fixed $L$, the set of kernel widths $\mathcal{L}$ can in theory
be chosen by minimizing $\epsilon$ in (\ref{eq:relaxed}) over $\mathcal{L}$
in addition to $w$. However, this results in a nonconvex optimization
problem since $w$ does not lie in the non-negative orthant. A parameter
search may not be practical as $\epsilon$ generally decreases as
the size and spread of $\mathcal{L}$ increases. However, a decrease
in $\epsilon$ does not always correspond to a decrease in MSE for
finite samples as high and low values of $h(l)$ can lead to inaccurate
density estimates. Thus we provide the following recommendations for
$\mathcal{L}$. Denote the value of the minimum value of $l$ so that
$\tilde{\mathbf{f}}_{i,h(l_{min})}(\mathbf{X}_{j})>0$ $\forall i=1,2$
as $l_{min}$ and the diameter of the support $\mathcal{S}$ as $D$.
To ensure the density estimates are bounded away from zero, we require
that $\min(\mathcal{L})\geq l_{min}$. The weights in $w_{0}$ are
generally largest for the smallest values of $\mathcal{L}$ (see Fig.~\ref{fig:weights})
so $\min(\mathcal{L})$ should also be sufficiently larger than $l_{min}$
to render an adequate estimate. Similarly, $\max(\mathcal{L})$ should
be sufficiently smaller than $D$ as high bandwidth values lead to
high bias. The remaining $\mathcal{L}$ values are chosen to be equally
spaced between $\min(\mathcal{L})$ and $\max(\mathcal{L})$. 

An efficient way to choose $l_{min}$ and $l_{max}$ is to select
integers $k_{min}$ and $k_{max}$ and compute the $k_{min}$ and
$k_{max}$ nearest neighbor distances of all the data points. The
bandwidths $h(l_{min})$ and $h(l_{max})$ can then be chosen to be
the maximum of these corresponding distances. The parameters $l_{min}$
and $l_{max}$ can then be computed from the expression $h(l)=lN^{-\frac{1}{2d}}$.
This choice ensures that a minimum of $k_{min}$ points are within
the kernel bandwidth for the density estimates at all points and that
a maximum of $k_{max}$ points are within the kernel bandwidth for
the density estimates at one of the points.

As $L$ increases, the similarity of bandwidth values $h(l)$ and
basis functions $\psi_{i,d}(l)$ increases, resulting in a negligible
decrease in the bias. Hence $L$ should be chosen large enough to
decrease the bias but small enough so that the $h(l)$ values are
sufficiently distinct (typically $30\leq L\leq60$). 

\subsection{Convergence Rates Validation: R\'{e}nyi-$\alpha$ Divergence}

\label{subsec:validate}To validate our theory, we estimated the R\'{e}nyi-$\alpha$
divergence integral between two truncated multivariate Gaussian distributions
with varying dimension and sample sizes. The densities have means
$\bar{\mu}_{1}=0.7*\bar{1}_{d}$, $\bar{\mu}_{2}=0.3*\bar{1}_{d}$
and covariance matrices $0.4*I_{d}$ where $\bar{1}_{d}$ is a $d$-dimensional
vector of ones, and $I_{d}$ is a $d\times d$ identity matrix. We
used $\alpha=0.5$ and restricted the Gaussians to the unit cube.

\begin{figure}
\centering

\includegraphics[width=0.5\textwidth]{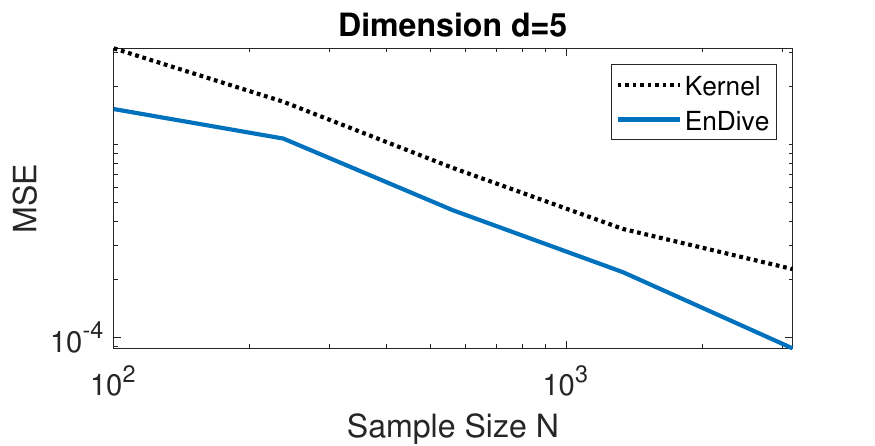}\includegraphics[width=0.5\textwidth]{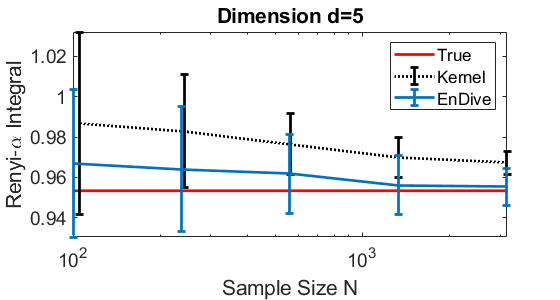}

\includegraphics[width=0.5\textwidth]{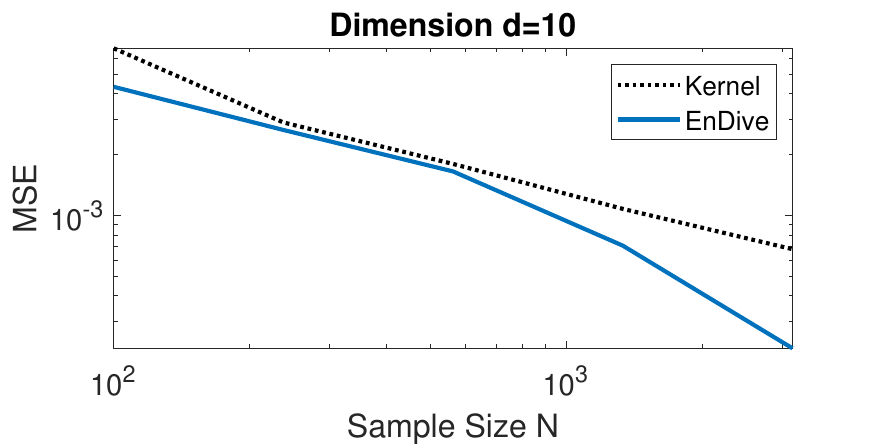}\includegraphics[width=0.5\textwidth]{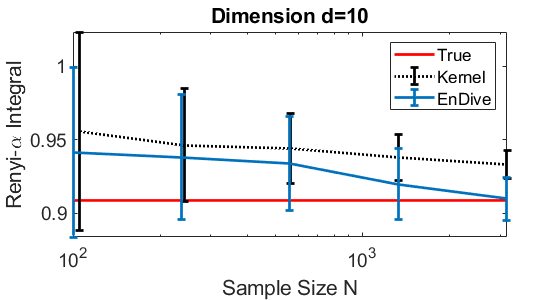}

\includegraphics[width=0.5\textwidth]{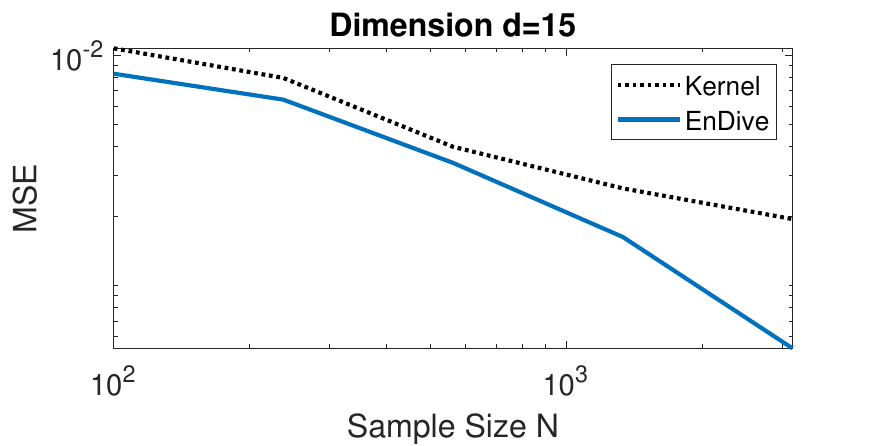}\includegraphics[width=0.5\textwidth]{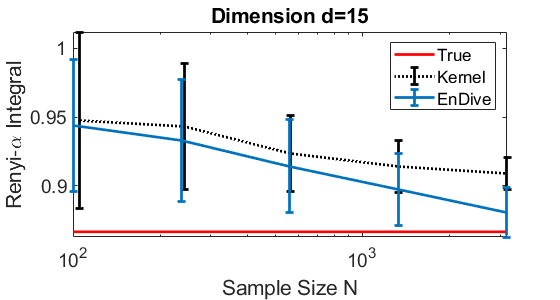}

\caption{\label{fig:mse}(Left) Log-log plot of MSE of the uniform kernel plug-in
(``Kernel'') and the optimally weighted EnDive estimator for various
dimensions and sample sizes. (Right) Plot of the average value of
the same estimators with standard error bars compared to the true
values being estimated. The proposed weighted ensemble estimator approaches
the theoretical rate (see Table~\ref{tab:slope}), performs better
than the plug-in estimator in terms of MSE, and is less biased.}
\end{figure}

The left plots in Figure~\ref{fig:mse} show the MSE (200 trials)
of the standard plug-in estimator implemented with a uniform kernel
and the proposed optimally weighted estimators EnDive for various
dimensions and sample sizes. The parameter set $\mathcal{L}$ was
selected based on a range of $k$-nearest neighbor distances. The
bandwidth used for the standard plug-in estimator was selected by
setting $h_{fixed}(l^{*})=l^{*}N^{-\frac{1}{d+1}}$ where $l^{*}$
was chosen from$\mathcal{L}$ to minimize the MSE of the plug-in estimator.
For all dimensions and sample sizes, EnDive outperforms the plug-in
estimator in terms of MSE. EnDive is also less biased than the plug-in
estimator and even has a lower variance at smaller sample sizes (e.g.
$N=100$). This reflects the strength of ensemble estimators: the
weighted sum of a set of relatively poor estimators can result in
a very good estimator. Note also that for the larger values of $N$
, the ensemble estimator MSE rates approach the theoretical rate based
on the estimated log-log slope given in Table~\ref{tab:slope}.

\begin{table}
\centering

\caption{\label{tab:slope}Negative log-log slope of the EnDive MSE as a function
of sample size for various dimensions. The slope is calculated beginning
at $N_{start}$. The negative slope is closer to 1 with $N_{start}=10^{2.375}$
than for $N_{start}=10^{2}$, indicating that the asymptotic rate
has not yet taken effect at $N_{start}=10^{2}$.}

\begin{tabular}{|c|c|c|c|}
\hline 
Estimator & $d=5$ & $d=10$ & $d=15$\tabularnewline
\hline 
\hline 
$N_{start}=10^{2}$ & $0.85$ & $0.84$ & $0.80$\tabularnewline
\hline 
$N_{start}=10^{2.375}$ & $0.96$ & $0.96$ & $0.95$\tabularnewline
\hline 
\end{tabular}
\end{table}

Our experiments indicated that the proposed ensemble estimator is
not sensitive to the tuning parameters. See~\cite{moon2016arxiv}
for more details.

\subsection{Central Limit Theorem Validation: KL Divergence}

To verify the central limit theorem of the EnDive estimator, we estimated
the KL divergence between two truncated Gaussian densities again restricted
to the unit cube. We conducted two experiments where 1) the densities
are different with means $\bar{\mu}_{1}=0.7*\bar{1}_{d}$, $\bar{\mu}_{2}=0.3*\bar{1}_{d}$
and covariance matrices $\sigma_{i}*I_{d}$, $\sigma_{1}=0.1$, $\sigma_{2}=0.3$;
and where 2) the densities are the same with means $0.3*\bar{1}_{d}$
and covariance matrices $0.3*I_{d}$. For both experiments, we chose
$d=6$ and four different sample sizes $N$. We found that the correspondence
between the quantiles of the standard normal distribution and the
quantiles of the centered and scaled EnDive estimator is very high
under all settings (see Table~\ref{tab:qq}), which validates Theorem~\ref{thm:clt}.

\begin{table}
\centering

\begin{tabular}{|c|c|c|c|c|}
\hline 
 & \multicolumn{2}{c|}{Same} & \multicolumn{2}{c|}{Different}\tabularnewline
\cline{2-5} 
$N$ & $1-\rho$ & $\beta$ & $1-\rho$ & $\beta$\tabularnewline
\hline 
\hline 
100 & $2.35\times10^{-4}$ & $1.014$ & $9.97\times10^{-4}$ & $0.993$\tabularnewline
\hline 
500 & $9.48\times10^{-5}$ & $1.007$ & $5.06\times10^{-4}$ & $0.999$\tabularnewline
\hline 
1000 & $8.27\times10^{-5}$ & $0.996$ & $4.30\times10^{-4}$ & $0.988$\tabularnewline
\hline 
5000 & $8.59\times10^{-5}$ & $0.995$ & $4.47\times10^{-4}$ & $1.005$\tabularnewline
\hline 
\end{tabular}\caption{Comparison between quantiles of a standard normal random variable
and the quantiles of the centered and scaled EnDive estimator applied
to the KL divergence when the distributions are the same and different.
Quantiles were computed from 10,000 trials. The parameter $\rho$
gives the correlation coefficient between the quantiles while $\beta$
is the estimated slope between the quantiles. The correspondence between
quantiles is very high for all cases.\label{tab:qq}}

\end{table}

\subsection{Bayes Error Rate Estimation on Single-Cell Data}

\begin{figure}
\includegraphics[width=0.5\textwidth]{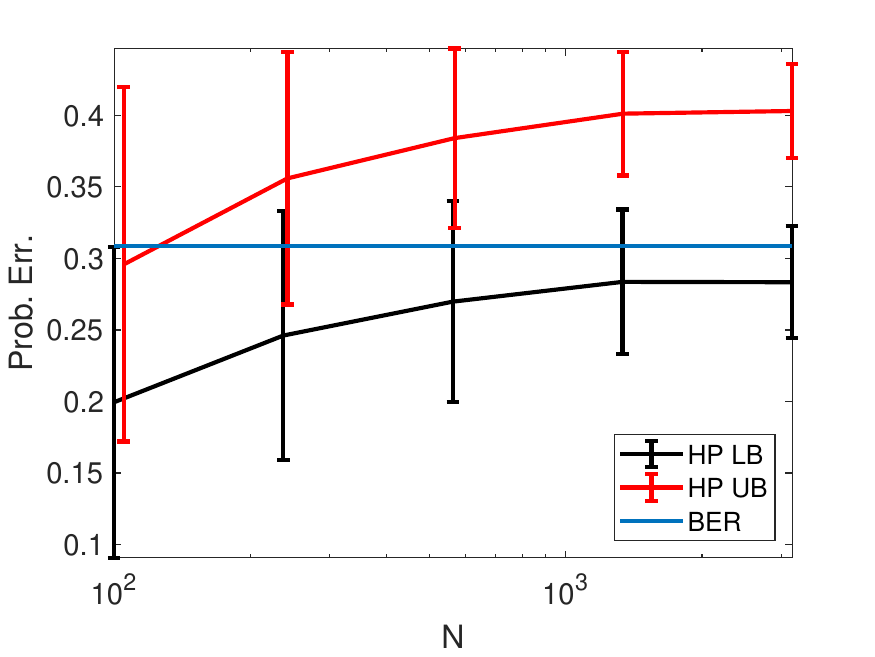}\includegraphics[width=0.5\textwidth]{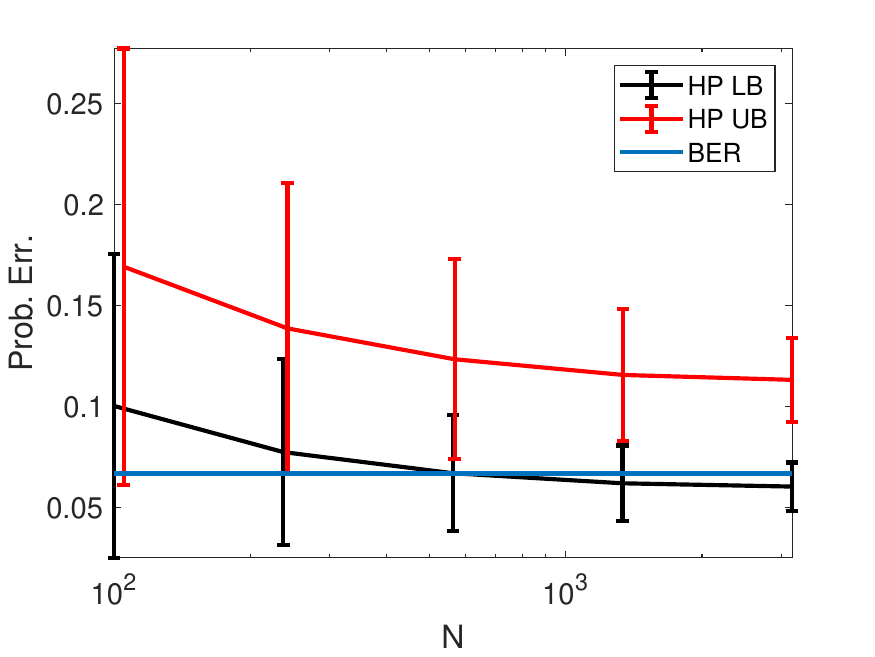}

\caption{Estimated upper (UB) and lower bounds (LB) on the BER based on estimating
the HP divergence between two 10-dimensional Gaussian distributions
with identity covariance matrix and a distance between means of 1
(left) and 3 (right), respectively. Estimates are calculated using
EnDive with error bars indicating the standard deviation from 400
trials. The upper bound is closer on average to the true BER when
$N$ is small ($\approx100-300$) and the distance between the means
is small. The lower bound is closer on average in all other cases.
\label{fig:BER}}

\end{figure}

Using the EnDive estimator, we estimate bounds on the Bayes error
rate (BER) of a classification problem involving MARS-seq single-cell
RNA-sequencing (scRNA-seq) data measured from developing mouse bone
marrow cells enriched for the myeloid and erythroid lineages~\cite{paul2015transcriptional}.
However, we first demonstrate the ability of EnDive to estimate bounds
on the BER of a simulated problem. In this simulation, the data are
drawn from two classes where each class distribution is a $d=10$
dimensional Gaussian distribution with different means and the identity
covariance matrix. We considered two cases: the distance between the
means is 1 and 3, respectively. The BER can be calculated in both
cases. We then estimated upper and lower bounds on the BER by estimating
the Henze-Penrose (HP) divergence~\cite{berisha2014bound,moon2015Bayes}.
Figure~\ref{fig:BER} shows the average estimated upper and lower
bounds on the BER with standard error bars for both cases. For all
tested sample sizes, the BER is within one standard deviation of the
estimated lower bound. The lower bound is also closer on average to
the BER for most of the tested sample sizes (lower sample sizes with
a smaller distance between means provide the exceptions). Generally,
these resuls indicate that the true BER is relatively close to the
estimated lower bound on average.

We now estimate similar bounds on the scRNA-seq classification problem
using EnDive. We consider the three most common cell types within
the data: erythrocytes, monocytes, and basophils ($N=1095,\,559,\,300,$
respectively). We estimate upper and lower bounds on the pairwise
BER between these classes using different combinations of genes selected
from the KEGG pathways associated with the hematopoietic cell lineage~\cite{kanehisa2000kegg,kanehisa2015kegg,kanehisa2016kegg}.
Each collection of genes contains 11-14 genes. The upper and lower
bounds on the BER are estimated using the Henze-Penrose divergence~\cite{berisha2014bound,moon2015Bayes}.
The standard deviation of the bounds for the KEGG-based genes is estimated
via 1000 bootstrap iterations. The KEGG-based bounds are compared
to BER bounds obtained from 1000 random selections of 12 genes. In
all cases, we compare the bounds to the performance of a quadratic
discriminant analysis classifier (QDA) with 10-fold cross validation.
Note that to correct for undersampling in scRNA-seq data, we first
impute the undersampled data using MAGIC~\cite{van2017magic}.

\begin{table}
\centering

\begin{tabular}{|l|c|c|c|c|c|}
\hline 
 & Platelets & Erythrocytes & Neutrophils & Macrophages & Random\tabularnewline
\hline 
\hline 
Erythrocytes vs. Monocytes, LB & $2.8\pm1.5$ & $1.2\pm0.6$ & $0.6\pm0.6$ & $8.5\pm1.2$ & $14.4\pm8.4$\tabularnewline
\hline 
Erythrocytes vs. Monocytes, UB & $5.3\pm2.9$ & $2.4\pm1.3$ & $1.2\pm1.3$ & $15.5\pm1.9$ & $23.2\pm12.3$\tabularnewline
\hline 
Erythrocytes vs. Monocytes, Prob. Error & $0.9$ & $0.4$ & $1.3$ & $3.4$ & $7.2\pm5.4$\tabularnewline
\hline 
Erythrocytes vs. Basophil, LB & $0.5\pm0.6$ & $0.05\pm0.12$ & $0.6\pm0.5$ & $5.1\pm0.9$ & $11.9\pm5.5$\tabularnewline
\hline 
Erythrocytes vs. Basophil, UB & $1.0\pm1.1$ & $0.1\pm0.2$ & $1.1\pm0.9$ & $9.6\pm1.6$ & $20.3\pm8.8$\tabularnewline
\hline 
Erythrocytes vs. Basophil, Prob. Error & $1.2$ & $0.3$ & $1.9$ & $3.6$ & $6.8\pm5.0$\tabularnewline
\hline 
Basophil vs. Monocytes, LB & $31.1\pm1.8$ & $27.8\pm3.1$ & $27.1\pm2.6$ & $31.6\pm1.3$ & $32.1\pm2.6$\tabularnewline
\hline 
Basophil vs. Monocytes, UB & $42.8\pm1.4$ & $39.9\pm2.8$ & $39.4\pm2.4$ & $43.2\pm1.0$ & $43.5\pm1.2$\tabularnewline
\hline 
Basophil vs. Monocytes, Prob. Error & $28.8$ & $30.9$ & $23.9$ & $22.4$ & $29.7\pm5.7$\tabularnewline
\hline 
\end{tabular}

\caption{Misclassification rate of a QDA classifier and estimated upper bounds
(UB) and lower bounds (LB) of the pairwise BER between mouse bone
marrow cell types using the Henze-Penrose divergence applied to different
combinations of genes selected from the KEGG pathways associated with
the hematopoietic cell lineage. Results are presented as percentages
in the form of mean$\pm$standard deviation. Based on these results,
erythrocytes are relatively easy to distinghuish from the other two
cell types using these gene sets. \label{tab:scRNAseq}}
\end{table}

All results are given in Table~\ref{tab:scRNAseq}. From these results,
we note that erythrocytes are relatively easy to distinguish from
the other two cell types as the BER lower bounds are within nearly
two standard deviations of zero when using genes associated with platelet,
erythrocyte, and neutrophil development as well as a random selection
of 12 genes. This is corroborated by the QDA cross-validated results
which are all within 2 standard deviations of either the upper or
lower bound for these gene sets. In contrast, the macrophage associated
genes seem to be less useful for distinguishing erythrocytes than
the other gene sets. 

We also find that basophils are difficult to distinguish from monocytes
using these gene sets. Assuming the relative abundance of each cell
type is representative of the population, a trivial upper bound on
the BER is $300/(300+559)\approx0.35$ which is between all of the
estimated lower and upper bounds. The QDA results are also relatively
high (and may be overfitting the data in some cases, based on the
estimated BER bounds), suggesting that different genes should be explored
for this classification problem.

\section{Conclusion}

We derived convergence rates for a kernel density plug-in estimator
of divergence functionals. We generalized the theory of optimally
weighted ensemble estimation and derived an ensemble divergence estimator
EnDive that achieves the parametric rate when the densities are more
than $d$ times differentiable. The estimator we derive applies to
general bounded density support sets and does not require knowledge
of the support which is a distinct advantage over other estimators.
We also derived the asymptotic distribution of the estimator, provided
some guidelines for tuning parameter selection, and validated the
convergence rates for the case of empirical estimation of the R\'enyi-$\alpha$
divergence. We then performed experiments to examine the estimator's
robustness to the choice of tuning parameters, validated the central
limit theorem for KL divergence estimation, and estimated bounds on
the Bayes error rate for a single cell classification problem. 

We note that based on the proof techniques employed in our work, our
weighted ensemble estimators are easily extended beyond divergence
estimation to more general distributional functionals which may be
integral functionals of any number of probability distributions. We
also show in Appendix~\ref{sec:Mod} that EnDive can be easily modified
to obtain an estimator that achieves the parametric rate when the
densities are more than $d/2$ times differentiable and the functional
$g$ has a specific form that includes the R\'{e}nyi and KL divergences.
Future work includes extending this modification to functionals with
more general forms. An important divergence of interest in this context
is the Henze-Penrose divergence we use to bound the Bayes error. Further
future work will focus on extending this work on divergence estimation
to $k$-nn based estimators where knowledge of the support is again
not required. This will improve the computational burden as $k$-nn
estimators require fewer computations than standard KDEs.

\appendices

\section{Modified EnDive}

\label{sec:Mod}If the functional $g$ has a specific form, we can
modify the EnDive estimator to obtain an estimator that achieves the
parametric rate when $s>d/2$. Specifically, we have the following: 

\begin{theorem}\emph{ Assume that assumptions }$\mathcal{A}.0-\mathcal{A}.5$\emph{
hold. Furthermore, if $g(x,y)$ }has \emph{$k,l$-th order mixed derivatives
$\frac{\partial^{k+l}g(x,y)}{\partial x^{k}\partial y^{l}}$ that
depend on $x,y$ only through $x^{\alpha}y^{\beta}$ for some $\alpha,\beta\in\mathbb{R}$,
then for any positive integer $\lambda\geq2$, the bias of $\gh$
is} 
\begin{eqnarray}
\bias\left[\gh\right] & = & \sum_{j=1}^{\left\lfloor s\right\rfloor }c_{10,j}h^{j}+\sum_{q=1}^{\lambda/2}\sum_{j=0}^{\left\lfloor s\right\rfloor }c_{11,q,j}\frac{h^{j}}{\left(Nh^{d}\right)^{q}}\nonumber \\
 &  & +O\left(h^{s}+\frac{1}{\left(Nh^{d}\right)^{\frac{\lambda}{2}}}\right).\label{eq:bias2}
\end{eqnarray}

\end{theorem}

Divergence functionals that satisfy the mixed derivatives condition
required for (\ref{eq:bias2}) include the KL divergence and the R\'{e}nyi-$\alpha$
divergence. Obtaining similar terms for other divergence functionals
requires us to separate the dependence on $h$ of the derivatives
of $g$ evaluated at $\ez\fth i(\mathbf{Z})$. This is left for future
work. See Appendix~\ref{sec:BiasProof} for details.

As compared to (\ref{eq:bias1}), there are many more terms in (\ref{eq:bias2}).
These terms enable us to modify the EnDive estimator to achieve the
parametric MSE convergence rate when $s>d/2$ for an appropriate choice
of bandwidths whereas the terms in (\ref{eq:bias1}) require $s\geq d$
to achieve the same rate. This is accomplished by letting $h(l)$
decrease at a faster rate as follows.

Let $\delta>0$ and $h(l)=lN^{\frac{-1}{d+\delta}}$ where $l\in\mathcal{L}$.
The bias of each estimator in the resulting ensemble has terms proportional
to $l^{j-dq}N^{-\frac{j+q}{d+\delta}}$ where $j,q\geq0$ and $j+q>0$.
Then the bias of $\tilde{\mathbf{G}}_{h(l)}$ satisfies condition
$\mathcal{C}.1$ if $\phi_{j,q,d}(N)=N^{-\frac{j+q}{d+\delta}}$,
$\psi_{j,q}(l)=l^{j-dq}$, and
\begin{equation}
J=\left\{ \{j,q\}:0<j+q<(d+1)/2,\,q\in\{0,1,2,\dots,\lambda/2\},\,j\in\{0,1,2,\dots,\left\lfloor s\right\rfloor \}\right\} ,\label{eq:J}
\end{equation}
as long as $L>|J|=I$. The variance also satisfies condition $\mathcal{C}.2$.
The optimal weight $w_{0}$ is found by using (\ref{eq:relaxed})
to obtain an optimally weighted plug-in divergence functional estimator
$\tilde{\mathbf{G}}_{w_{0}}$ that achieves the parametric convergence
rate if $\lambda\geq d/\delta+1$ and if $s\geq(d+\delta)/2$. Otherwise,
if $s<(d+\delta)/2$ we can only guarantee the MSE rate up to $O\left(\frac{1}{N^{2s/(d+\delta)}}\right)$.
We refer to this estimator as the modified EnDive estimator and denote
it as $\tilde{\mathbf{G}}_{\text{Mod}}$. The ensemble estimator $\tilde{\mathbf{G}}_{\text{Mod}}$
is summarized in Algorithm~\ref{alg:estimator} when $\delta=1$. 

\begin{algorithm*}
\begin{algorithmic}[1]
\renewcommand{\algorithmicrequire}{\textbf{Input:}} \renewcommand{\algorithmicensure}{\textbf{Output:}}

\REQUIRE $\eta$, $L$ positive real numbers $\mathcal{L}$, samples
$\left\{ \mathbf{Y}_{1},\dots,\mathbf{Y}_{N}\right\} $ from $f_{1}$,
samples $\left\{ \mathbf{X}_{1},\dots,\mathbf{X}_{N}\right\} $ from
$f_{2}$, dimension $d$, function $g$, kernel $K$

\ENSURE The modified EnDive estimator $\tilde{\mathbf{G}}_{\text{Mod}}$

\STATE Solve for $w_{0}$ using (\ref{eq:relaxed}) with $\phi_{j,q,d}(N)=N^{-\frac{j+q}{d+1}}$
and basis functions $\psi_{j,q}(l)=l^{j-dq}$, $l\in\bar{l}$, and
$\{i,j\}\in J$ defined in (\ref{eq:J})

\FORALL{$l\in\bar{l}$}

\STATE $h(l)\leftarrow lN^{\frac{-1}{d+1}}$

\FOR{$i=1$ to $N$}

\STATE $\ftl 1(\mathbf{X}_{i})\leftarrow\frac{1}{Nh(l)^{d}}\sum_{j=1}^{N}K\left(\frac{\mathbf{X}_{i}-\mathbf{Y}_{j}}{h(l)}\right)$,
$\ftl 2(\mathbf{X}_{i})\leftarrow\frac{1}{(N-1)h(l)^{d}}\sum_{\substack{j=1\\
j\neq i
}
}^{N}K\left(\frac{\mathbf{X}_{i}-\mathbf{X}_{j}}{h(l)}\right)$

\ENDFOR 

\STATE $\tilde{\mathbf{G}}_{h(l)}\leftarrow\frac{1}{N}\sum_{i=1}^{N}g\left(\ftl 1(\mathbf{X}_{i}),\ftl 2(\mathbf{X}_{i})\right)$

\ENDFOR 

\STATE $\tilde{\mathbf{G}}_{\text{Mod}}\leftarrow\sum_{l\in\mathcal{L}}w_{0}(l)\tilde{\mathbf{G}}_{h(l)}$

\end{algorithmic}

\caption{The modified EnDive estimator \label{alg:estimator}}
\end{algorithm*}

The parametric rate can be achieved with $\tilde{\mathbf{G}}_{\text{Mod}}$
under less strict assumptions on the smoothness of the densities than
those required for $\tilde{\mathbf{G}}_{\text{EnDive}}$. Since $\delta>0$
can be arbitrary, it is theoretically possible to achieve the parametric
rate with the modified estimator as long as $s>d/2$. This is consistent
with the rate achieved by the more complex estimators proposed in~\cite{krishnamurthy2014divergence}.
We also note that the central limit theorem applies and that the convergence
is uniform as Theorem~\ref{thm:uniform} applies for $s>\left\lfloor (d+\delta)/2\right\rfloor $
and $s\geq(d+\delta)/2$.

These rate improvements come at a cost in the number of parameters
$L$ required to implement the weighted ensemble estimator. If $s\geq\frac{d+\delta}{2}$
then the size of $J$ for $\tilde{\mathbf{G}}_{\text{Mod}}$ is on
the order of $\frac{d^{2}}{8\delta}$. This may lead to increased
variance of the ensemble estimator as indicated by (\ref{eq:ens_var}). 

So far $\tilde{\mathbf{G}}_{\text{Mod}}$ can only be applied to functionals
$g(x,y)$ with mixed derivatives of the form of $x^{\alpha}y^{\beta}$.
Future work is required to extend this estimator to other functionals
of interest.

\section{General Results}

\label{sec:general}Here we present the generalized forms of Theorems~\ref{thm:bias}
and \ref{thm:variance} where the sample sizes and bandwidths of the
two datasets are allowed to differ. In this case, the KDEs are
\begin{eqnarray*}
\ft 1(\mathbf{X}_{j}) & = & \frac{1}{N_{1}h_{1}^{d}}\sum_{i=1}^{N_{1}}K\left(\frac{\mathbf{X}_{j}-\mathbf{Y}_{i}}{h_{1}}\right),\\
\ft 2(\mathbf{X}_{j}) & = & \frac{1}{M_{2}h_{2}^{d}}\sum_{\substack{i=1\\
i\neq j
}
}^{N_{2}}K\left(\frac{\mathbf{X}_{j}-\mathbf{X}_{i}}{h_{2}}\right),
\end{eqnarray*}
where $M_{2}=N_{2}-1$. $G\left(f_{1},f_{2}\right)$ is then approximated
as 
\begin{equation}
\gt=\frac{1}{N_{2}}\sum_{i=1}^{N_{2}}g\left(\ft 1\left(\mathbf{X}_{i}\right),\ft 2\left(\mathbf{X}_{i}\right)\right).\label{eq:estimator-1}
\end{equation}

We also generalize the bias result to the case where the kernel $K$
has order $\nu$ which means that the $j$th moment of the kernel
$K_{i}$ defined as $\int t^{j}K_{i}(t)dt$ is zero for all $j=1,\dots,\nu-1$
and $i=1,\dots,d$ where $K_{i}$ is the kernel in the $i$th coordinate.
Note that symmetric product kernels have order $\nu\geq2$. The following
theorem on the bias follows under assumptions $\mathcal{A}.0-\mathcal{A}.5$:

\begin{theorem}\label{thm:bias-1}\emph{For general $g$, the bias
of the plug-in estimator $\gt$ is of the form 
\begin{eqnarray}
\bias\left[\gt\right] & = & \sum_{j=1}^{r}\left(c_{4,1,j}h_{1}^{j}+c_{4,2,j}h_{2}^{j}\right)+\sum_{j=1}^{r}\sum_{i=1}^{r}c_{5,i,j}h_{1}^{j}h_{2}^{i}+O\left(h_{1}^{s}+h_{2}^{s}\right)\nonumber \\
 &  & +c_{9,1}\frac{1}{N_{1}h_{1}^{d}}+c_{9,2}\frac{1}{N_{2}h_{2}^{d}}+o\left(\frac{1}{N_{1}h_{1}^{d}}+\frac{1}{N_{2}h_{2}^{d}}\right).\label{eq:bias1-1}
\end{eqnarray}
Furthermore, if $g(x,y)$ }has \emph{$k,l$-th order mixed derivatives
$\frac{\partial^{k+l}g(x,y)}{\partial x^{k}\partial y^{l}}$ that
depend on $x,y$ only through $x^{\alpha}y^{\beta}$ for some $\alpha,\beta\in\mathbb{R}$,
then for any positive integer $\lambda\geq2$, the bias is of the
form}
\begin{eqnarray}
\bias\left[\gt\right] & = & \sum_{j=1}^{r}\left(c_{4,1,j}h_{1}^{j}+c_{4,2,j}h_{2}^{j}\right)+\sum_{j=1}^{r}\sum_{i=1}^{r}c_{5,i,j}h_{1}^{j}h_{2}^{i}+O\left(h_{1}^{s}+h_{2}^{s}\right)\nonumber \\
 &  & \sum_{j=1}^{\lambda/2}\sum_{m=0}^{r}\left(c_{9,1,j,m}\frac{h_{1}^{m}}{\left(N_{1}h_{1}^{d}\right)^{j}}+c_{9,2,j,m}\frac{h_{2}^{m}}{\left(N_{2}h_{2}^{d}\right)^{j}}\right)\nonumber \\
 &  & +\sum_{j=1}^{\lambda/2}\sum_{m=0}^{r}\sum_{i=1}^{\lambda/2}\sum_{n=0}^{r}c_{9,j,i,m,n}\frac{h_{1}^{m}h_{2}^{n}}{\left(N_{1}h_{1}^{d}\right)^{j}\left(N_{2}h_{2}^{d}\right)^{i}}\nonumber \\
 &  & +O\left(\frac{1}{\left(N_{1}h_{1}^{d}\right)^{\frac{\lambda}{2}}}+\frac{1}{\left(N_{2}h_{2}^{d}\right)^{\frac{\lambda}{2}}}\right).\label{eq:bias2-1}
\end{eqnarray}

\end{theorem}

Note that the bandwidth and sample size terms do not depend on the
order of the kernel $\nu$. Thus using a higher-order kernel does
not provide any benefit in the convergence rates. This lack of improvement
is due to the bias of the density estimators at the boundary of the
densities' support sets. To obtain better convergence rates using
higher-order kernels, boundary correction would be necessary~\cite{kandasamy2015nonparametric,krishnamurthy2014divergence}.
In contrast, we improve the convergence rates by using a weighted
ensemble that does not require boundary correction.

The variance result requires much less strict assumptions than the
bias results:

\begin{theorem} \emph{\label{thm:var-1}Assume that the functional}
\emph{$g$ in~}(\ref{eq:fdiv})\emph{ is Lipschitz continuous in
both of its arguments with Lipschitz constant $C_{g}$. Then the variance
of the plug-in estimator $\gt$ is bounded by
\[
\var\left[\gt\right]\leq C_{g}^{2}||K||_{\infty}^{2}\left(\frac{10}{N_{2}}+\frac{N_{1}}{N_{2}^{2}}\right).
\]
}

\end{theorem}

The proofs of these theorems are in Appendices~\ref{sec:BiasProof}
and \ref{sec:VarProof}. Theorems~\ref{thm:bias} and \ref{thm:variance}
then follow.

\section{Proof of Theorem~\ref{thm:boundary} (Boundary Conditions)}

\label{sec:boundaryProof}Consider a uniform rectangular kernel $K(x)$
that satisfies $K(x)=1$ for all $x$ such that $||x||_{1}\leq1/2$.
Also consider the family of probability densities $f$ with rectangular
support $\mathcal{S}=[-1,1]^{d}$. We will prove Theorem~\ref{thm:boundary}
which is that that $\S$ satisfies the following smoothness condition
$(\mathcal{A}.5$): for any polynomial $p_{x}(u):\mathbb{R}^{d}\rightarrow\mathbb{R}$
of order $q\leq r=\left\lfloor s\right\rfloor $ with coefficients
that are $r-q$ times differentiable wrt $x$, 
\begin{equation}
\int_{x\in\S}\left(\int_{u:||u||_{1}\leq\frac{1}{2},\,x+uh\notin\S}p_{x}(u)du\right)^{t}dx=v_{t}(h),\label{eq:smoothness}
\end{equation}
where $v_{t}(h)$ has the expansion 
\[
v_{t}(h)=\sum_{i=1}^{r-q}e_{i,q,t}h^{i}+o\left(h^{r-q}\right).
\]
Note that the inner integral forces the $x$'s under consideration
to be boundary points via the constraint $x+uh\notin\S$.

\subsection{Single Coordinate Boundary Point}

We begin by focusing on points $x$ that are boundary points by virtue
of a single coordinate $x_{i}$ such that $x_{i}+u_{i}h\notin\S$.
Without loss of generality, assume that $x_{i}+u_{i}h>1$. The inner
integral in (\ref{eq:smoothness}) can then be evaluated first wrt
all coordinates other than $i$. Since all of these coordinates lie
within the support, the inner integral over these coordinates will
amount to integration of the polynomial $p_{x}(u)$ over a symmetric
$d-1$ dimensional rectangular region $\left|u_{j}\right|\leq\frac{1}{2}$
for all $j\neq i$. This yields a function $\sum_{m=1}^{q}\tilde{p}_{m}(x)u_{i}^{m}$
where the coefficients $\tilde{p}_{m}(x)$ are each $r-q$ times differentiable
wrt $x$.

With respect to the $u_{i}$ coordinate, the inner integral will have
limits from $\frac{1-x_{i}}{h}$ to $\frac{1}{2}$ for some $1>x_{i}>1-\frac{h}{2}$.
Consider the $\tilde{p}_{q}(x)u_{i}^{q}$ monomial term. The inner
integral wrt this term yields 
\begin{equation}
\sum_{m=1}^{q}\tilde{p}_{m}(x)\int_{\frac{1-x_{i}}{h}}^{\frac{1}{2}}u_{i}^{m}du_{i}=\sum_{m=1}^{q}\tilde{p}_{m}(x)\frac{1}{m+1}\left(\frac{1}{2^{m+1}}-\left(\frac{1-x_{i}}{h}\right)^{m+1}\right).\label{eq:poly1}
\end{equation}
Raising the right hand side of (\ref{eq:poly1}) to the power of $t$
results in an expression of the form 
\begin{equation}
\sum_{j=0}^{qt}\check{p}_{j}(x)\left(\frac{1-x_{i}}{h}\right)^{j},\label{eq:poly2}
\end{equation}
where the coefficients $\check{p}_{j}(x)$ are $r-q$ times differentiable
wrt $x$. Integrating (\ref{eq:poly2}) over all the coordinates in
$x$ other than $x_{i}$ results in an expression of the form 
\begin{equation}
\sum_{j=0}^{qt}\bar{p}_{j}(x_{i})\left(\frac{1-x_{i}}{h}\right)^{j},\label{eq:poly3}
\end{equation}
 where again the coefficients$\bar{p}_{j}(x_{i})$ are $r-q$ times
differentiable wrt $x_{i}$. Note that since the other cooordinates
of $x$ other than $x_{i}$ are far away from the boundary, the coefficients
$\bar{p}_{j}(x_{i})$ are independent of $h$. To evaluate the integral
of (\ref{eq:poly3}), consider the $r-q$ term Taylor series expansion
of $\bar{p}_{j}(x_{i})$ around $x_{i}=1$. This will yield terms
of the form 
\begin{eqnarray*}
\int_{1-h/2}^{1}\frac{\left(1-x_{i}\right)^{j+k}}{h^{k}}dx_{i} & = & \left.-\frac{\left(1-x_{i}\right)^{j+k+1}}{h^{k}(j+k+1)}\right|_{x_{i}=1-h/2}^{x_{i}=1}\\
 & = & \frac{h^{j+1}}{(j+k+1)2^{j+k+1}},
\end{eqnarray*}
for $0\leq j\leq r-q$, and $0\leq k\leq qt$. Combining terms results
in the expansion $v_{t}(h)=\sum_{i=1}^{r-q}e_{i,q,t}h^{i}+o\left(h^{r-q}\right)$.

\subsection{Multiple Coordinate Boundary Point}

The case where multiple coordinates of the point $x$ are near the
boundary is a straightforward extension of the single boundary point
case so we only sketch the main ideas here. As an example, consider
the case where 2 of the coordinates are near the boundary. Assume
for notational ease that they are $x_{1}$ and $x_{2}$ and that $x_{1}+u_{1}h>1$
and $x_{2}+u_{2}h>1$. The inner integral in (\ref{eq:smoothness})
can again be evaluated first wrt all coordinates other than 1 and
2. This yields a function $\sum_{m,j=1}^{q}\tilde{p}_{m,j}(x)u_{1}^{m}u_{2}^{j}$
where the coefficients $\tilde{p}_{m,j}(x)$ are each $r-q$ times
differentiable wrt $x$. Integrating this wrt $x_{1}$ and $x_{2}$
and then raising the result to the power of $t$ yields a double sum
similar to (\ref{eq:poly2}). Integrating this over all the coordinates
in $x$ other than $x_{1}$ and $x_{2}$ gives a double sum similar
to (\ref{eq:poly3}). Then a Taylor series expansion of the coefficients
and integration over $x_{1}$ and $x_{2}$ yields the result.

\section{Proof of Theorem~\ref{thm:bias-1} (Bias)}

\label{sec:BiasProof} In this appendix, we prove the bias results
in Thm.~\ref{thm:bias-1}. The bias of the base kernel density plug-in
estimator $\gt$ can be expressed as 
\begin{eqnarray}
\bias\left[\gt\right] & = & \mathbb{E}\left[g\left(\ft 1(\mathbf{Z}),\ft 2(\mathbf{Z})\right)-g\left(f_{1}(\mathbf{Z}),f_{2}(\mathbf{Z})\right)\right]\nonumber \\
 & = & \mathbb{E}\left[g\left(\ft 1(\mathbf{Z}),\ft 2(\mathbf{Z})\right)-g\left(\ez\ft 1(\mathbf{Z}),\ez\ft 2(\mathbf{Z})\right)\right]\nonumber \\
 &  & +\mathbb{E}\left[g\left(\ez\ft 1(\mathbf{Z}),\ez\ft 2(\mathbf{Z})\right)-g\left(f_{1}(\mathbf{Z}),f_{2}(\mathbf{Z})\right)\right],\label{eq:gsplit}
\end{eqnarray}
where $\mathbf{Z}$ is drawn from $f_{2}$. The first term is the
``variance'' term while the second is the ``bias'' term. We bound
these terms using Taylor series expansions under the assumption that
$g$ is infinitely differentiable. The Taylor series expansion of
the variance term in (\ref{eq:gsplit}) will depend on variance-like
terms of the KDEs while the Taylor series expansion of the bias term
in (\ref{eq:gsplit}) will depend on the bias of the KDEs.

The Taylor series expansion of $g\left(\ez\ft 1(\mathbf{Z}),\ez\ft 2(\mathbf{Z})\right)$
around $f_{1}(\mathbf{Z})$ and $f_{2}(\mathbf{Z})$ is
\begin{eqnarray}
g\left(\ez\ft 1(\mathbf{Z}),\ez\ft 2(\mathbf{Z})\right) & = & \sum_{i=0}^{\infty}\sum_{j=0}^{\infty}\left(\left.\frac{\partial^{i+j}g(x,y)}{\partial x^{i}\partial y^{j}}\right|_{\substack{x=f_{1}(\mathbf{Z})\\
y=f_{2}(\mathbf{Z})
}
}\right)\frac{\bias_{\mathbf{Z}}^{i}\left[\ft 1(\mathbf{Z})\right]\bias_{\mathbf{Z}}^{j}\left[\ft 2(\mathbf{Z})\right]}{i!j!}\label{eq:gtaylor1}
\end{eqnarray}
where $\bias_{\mathbf{Z}}^{j}\left[\ft i(\mathbf{Z})\right]=\left(\ez\ft i(\mathbf{Z})-f_{i}(\mathbf{Z})\right)^{j}$
is the bias of $\ft i$ at the point $\mathbf{Z}$ raised to the power
of $j$. This expansion can be used to control the second term (the
bias term) in (\ref{eq:gsplit}). To accomplish this, we require an
expression for $\ez\ft i(\mathbf{Z})-f_{i}(\mathbf{Z})=\bias_{\mathbf{Z}}\left[\ft i(\mathbf{Z})\right]$.

To obtain an expression for $\bias_{\mathbf{Z}}\left[\ft i(\mathbf{Z})\right]$,
we consider separately the cases when $\mathbf{Z}$ is in the interior
of the support $\mathcal{S}$ or when $\mathbf{Z}$ is near the boundary
of the support. A point $X\in\mathcal{S}$ is defined to be in the
interior of $\mathcal{S}$ if for all $Y\notin\mathcal{S}$, $K\left(\frac{X-Y}{h_{i}}\right)=0$.
A point $X\in\mathcal{S}$ is near the boundary of the support if
it is not in the interior. Denote the region in the interior and near
the boundary wrt $h_{i}$ as $\mathcal{S}_{I_{i}}$ and $\mathcal{S}_{B_{i}}$,
respectively. We will need the following.

\begin{lemma}\label{lem:interior}Let $\mathbf{Z}$ be a realization
of the density $f_{2}$ independent of $\ft i$ for $i=1,2$. Assume
that the densities $f_{1}$ and $f_{2}$ belong to $\Sigma(s,L)$.
Then for $\mathbf{Z}\in\mathcal{S}_{I_{i}}$, 
\begin{equation}
\mathbb{E}_{\mathbf{Z}}\left[\ft i(\mathbf{Z})\right]=f_{i}(\mathbf{Z})+\sum_{j=\nu/2}^{\left\lfloor s/2\right\rfloor }c_{i,j}(\mathbf{Z})h_{i}^{2j}+O\left(h_{i}^{s}\right).\label{eq:interior}
\end{equation}

\end{lemma}
\begin{IEEEproof}
Obtaining the lower order terms in (\ref{eq:interior}) is a common
result in kernel density estimation. However, since we also require
the higher order terms, we present the proof here. Additionally, some
of the results in this proof will be useful later. From the linearity
of the KDE, we have that if $\mathbf{X}$ is drawn from $f_{i}$ and
is independent of $\mathbf{Z}$, then 
\begin{eqnarray}
\mathbb{E}_{\mathbf{Z}}\ft i(\mathbf{Z}) & = & \ez\left[\frac{1}{h_{i}^{d}}K\left(\frac{\mathbf{X}-\mathbf{Z}}{h_{i}}\right)\right]\nonumber \\
 & = & \int\frac{1}{h_{i}^{d}}K\left(\frac{x-\mathbf{Z}}{h_{i}}\right)f_{i}(x)dx\nonumber \\
 & = & \int K\left(t\right)f_{i}(th_{i}+\mathbf{Z})dt,\label{eq:substitution}
\end{eqnarray}
where the last step follows from the substitution $t=\frac{x-\mathbf{Z}}{h_{i}}$.
Since the density $f_{i}$ belongs to $\Sigma(s,K)$, using multi-index
notation we can expand it as 
\begin{equation}
f_{i}(th_{i}+\mathbf{Z})=f_{i}(\mathbf{Z})+\sum_{0<|\alpha|\leq\left\lfloor s\right\rfloor }\frac{D^{\alpha}f_{i}(\mathbf{Z})}{\alpha!}\left(th_{i}\right)^{\alpha}+O\left(\left\Vert th_{i}\right\Vert ^{s}\right),\label{eq:fTaylor}
\end{equation}
where $\alpha!=\alpha_{1}!\alpha_{2}!\dots\alpha_{d}!$ and $Y^{\alpha}=Y_{1}^{\alpha_{1}}Y_{2}^{\alpha_{2}}\dots Y_{d}^{\alpha_{d}}$.
Combining (\ref{eq:substitution}) and (\ref{eq:fTaylor}) gives 
\begin{eqnarray*}
\mathbb{E}_{\mathbf{Z}}\ft i(\mathbf{Z}) & = & f_{i}(\mathbf{Z})+\sum_{0<|\alpha|\leq\left\lfloor s\right\rfloor }\frac{D^{\alpha}f_{i}(\mathbf{Z})}{\alpha!}h_{i}^{|\alpha|}\int t^{\alpha}K(t)dt+O(h_{i}^{s})\\
 & = & f_{i}(\mathbf{Z})+\sum_{j=\nu/2}^{\left\lfloor s/2\right\rfloor }c_{i,j}(\mathbf{Z})h_{i}^{2j}+O(h_{i}^{s}),
\end{eqnarray*}
where the last step follows from the fact that $K$ is symmetric and
of order $\nu$.
\end{IEEEproof}
To obtain a similar result for the case when $\mathbf{Z}$ is near
the boundary of $\mathcal{S}$, we use assumption $\mathcal{A}.5$.

\begin{lemma}\label{lem:boundary}\emph{Let $\gamma(x,y)$ be an
arbitrary function satisfying $\sup_{x,y}|\gamma(x,y)|<\infty$. Let
$\mathcal{S}$ satisfy the boundary smoothness conditions of Assumption
$\mathcal{A}.5$. Assume that the densities $f_{1}$ and $f_{2}$
belong to $\Sigma(s,L)$ and let $\mathbf{Z}$ be a realization of
the density $f_{2}$ independent of $\ft i$ for $i=1,2$. Let $h^{'}=\min\left(h_{1},h_{2}\right)$.
Then 
\begin{eqnarray}
\bE\left[1_{\left\{ \mathbf{Z}\in\mathcal{S}_{B_{i}}\right\} }\gamma\left(f_{1}(\mathbf{Z}),f_{2}(\mathbf{Z})\right)\bias_{\mathbf{Z}}^{t}\left[\ft i(\mathbf{Z})\right]\right] & = & \sum_{j=1}^{r}c_{4,i,j,t}h_{i}^{j}+o\left(h_{i}^{r}\right)\label{eq:boundary_single}\\
\bE\left[1_{\left\{ \mathbf{Z}\in\mathcal{S}_{B_{1}}\cap\mathcal{S}_{B_{2}}\right\} }\gamma\left(f_{1}(\mathbf{Z}),f_{2}(\mathbf{Z})\right)\bias_{\mathbf{Z}}^{t}\left[\ft 1(\mathbf{Z})\right]\bias_{\mathbf{Z}}^{q}\left[\ft 2(\mathbf{Z})\right]\right] & = & \sum_{j=0}^{r-1}\sum_{i=0}^{r-1}c_{4,j,i,q,t}h_{1}^{j}h_{2}^{i}h^{'}+o\left(\left(h^{'}\right)^{r}\right)\label{eq:boundary_both}
\end{eqnarray}
}

\end{lemma}
\begin{IEEEproof}
For fixed $X$ near the boundary of $\mathcal{S}$, we have 
\begin{eqnarray*}
\mathbb{E}\left[\ft i(X)\right]-f_{i}(X) & = & \frac{1}{h_{i}^{d}}\int_{Y:Y\in\mathcal{S}}K\left(\frac{X-Y}{h_{i}}\right)f_{i}(Y)dY-f_{i}(X)\\
 & = & \left[\frac{1}{h_{i}^{d}}\int_{Y:K\left(\frac{X-Y}{h_{i}}\right)>0}K\left(\frac{X-Y}{h_{i}}\right)f_{i}(Y)dY-f_{i}(X)\right]\\
 &  & -\left[\frac{1}{h_{i}^{d}}\int_{Y:Y\notin\mathcal{S}}K\left(\frac{X-Y}{h_{i}}\right)f_{i}(Y)dY\right]\\
 & = & T_{1,i}(X)-T_{2,i}(X).
\end{eqnarray*}

Note that in $T_{1,i}(X)$, we are extending the integral beyond the
support of the density $f_{i}$. However, by using the same Taylor
series expansion method as in the proof of Lemma~\ref{lem:interior},
we always evaluate $f_{i}$ and its derivatives at the point $X$
which is within the support of $f_{i}$. Thus it does not matter how
we define an extension of $f_{i}$ since the Taylor series will remain
the same. Thus $T_{1,i}(X)$ results in an identical expression to
that obtained from (\ref{eq:interior}).

For the $T_{2,i}(X)$ term, we expand it as follows using multi-index
notation as 
\begin{eqnarray*}
T_{2,i}(X) & = & \frac{1}{h_{i}^{d}}\int_{Y:Y\notin\mathcal{S}}K\left(\frac{X-Y}{h_{i}}\right)f_{i}(Y)dY\\
 & = & \int_{u:h_{i}u+X\notin\mathcal{S},K(u)>0}K\left(u\right)f_{i}(X+h_{i}u)du\\
 & = & \sum_{|\alpha|\leq r}\frac{h_{i}^{|\alpha|}}{\alpha!}\int_{u:h_{i}u+X\notin\mathcal{S},K(u)>0}K\left(u\right)D^{\alpha}f_{i}(X)u^{\alpha}du+o\left(h_{i}^{r}\right).
\end{eqnarray*}
Recognizing that the $|\alpha|$th derivative of $f_{i}$ is $r-|\alpha|$
times differentiable, we can apply assumption $\mathcal{A}.5$ to
obtain the expectation of $T_{2,i}(X)$ wrt $X$: 
\begin{eqnarray*}
\bE\left[T_{2,i}(\mathbf{X})\right] & = & \frac{1}{h_{i}^{d}}\int_{X}\int_{Y:Y\notin\mathcal{S}}K\left(\frac{X-Y}{h_{i}}\right)f_{i}(Y)dYf_{2}(X)dx\\
 & = & \sum_{|\alpha|\leq r}\frac{h_{i}^{|\alpha|}}{\alpha!}\int_{X}\int_{u:h_{i}u+X\notin\mathcal{S},K(u)>0}K\left(u\right)D^{\alpha}f_{i}(X)u^{\alpha}duf_{2}(X)dX+o\left(h_{i}^{r}\right)\\
 & = & \sum_{|\alpha|\leq r}\frac{h_{i}^{|\alpha|}}{\alpha!}\left[\sum_{1\leq|\beta|\leq r-|\alpha|}e_{\beta,r-|\alpha|}h_{i}^{|\beta|}+o\left(h_{i}^{r-|\alpha|}\right)\right]+o\left(h_{i}^{r}\right)\\
 & = & \sum_{j=1}^{r}e_{j}h_{i}^{j}+o\left(h_{i}^{r}\right).
\end{eqnarray*}

Similarly, we find that 
\begin{eqnarray*}
\bE\left[\left(T_{2,i}(\mathbf{X})\right)^{t}\right] & = & \frac{1}{h_{i}^{dt}}\int_{X}\left(\int_{Y:Y\notin\mathcal{S}}K\left(\frac{X-Y}{h_{i}}\right)f_{i}(Y)dY\right)^{t}f_{2}(X)dx\\
 & = & \int_{X}\left(\sum_{|\alpha|\leq r}\frac{h_{i}^{|\alpha|}}{\alpha!}\int_{u:h_{i}u+X\notin\mathcal{S},K(u)>0}K\left(u\right)D^{\alpha}f_{i}(X)u^{\alpha}du\right)^{t}f_{2}(X)dX\\
 & = & \sum_{j=1}^{r}e_{j,t}h_{i}^{j}+o\left(h_{i}^{r}\right).
\end{eqnarray*}

Combining these results gives 
\begin{eqnarray*}
\bE\left[1_{\left\{ \mathbf{Z}\in\mathcal{S}_{B}\right\} }\gamma\left(f_{1}(\mathbf{Z}),f_{2}(\mathbf{Z})\right)\left(\ez\left[\ft i(\mathbf{Z})\right]-f_{i}(\mathbf{Z})\right)^{t}\right] & = & \bE\left[\gamma\left(f_{1}(\mathbf{Z}),f_{2}(\mathbf{Z})\right)\left(T_{1,i}(\mathbf{Z})-T_{2,i}(\mathbf{Z})\right)^{t}\right]\\
 & = & \bE\left[\gamma\left(f_{1}(\mathbf{Z}),f_{2}(\mathbf{Z})\right)\sum_{j=0}^{t}\binom{t}{j}\left(T_{1,i}(\mathbf{Z})\right)^{j}\left(-T_{2,i}(\mathbf{Z})\right)^{t-j}\right]\\
 & = & \sum_{j=1}^{r}c_{4,i,j,t}h_{i}^{j}+o\left(h_{i}^{r}\right),
\end{eqnarray*}
where the constants are functionals of the kernel, $\gamma$, and
the densities.

The expression in~(\ref{eq:boundary_both}) can be proved in a similar
manner. 
\end{IEEEproof}
Applying Lemmas~\ref{lem:interior} and \ref{lem:boundary} to (\ref{eq:gtaylor1})
gives 
\begin{equation}
\mathbb{E}\left[g\left(\ez\ft 1(\mathbf{Z}),\ez\ft 2(\mathbf{Z})\right)-g\left(f_{1}(\mathbf{Z}),f_{2}(\mathbf{Z})\right)\right]=\sum_{j=1}^{r}\left(c_{4,1,j}h_{1}^{j}+c_{4,2,j}h_{2}^{j}\right)+\sum_{j=0}^{r-1}\sum_{i=0}^{r-1}c_{5,i,j}h_{1}^{j}h_{2}^{i}h^{'}+o\left(h_{1}^{r}+h_{2}^{r}\right).\label{eq:biasresult1}
\end{equation}

For the variance term (the first term) in (\ref{eq:gsplit}), the
truncated Taylor series expansion of $g\left(\ft 1(\mathbf{Z}),\ft 2(\mathbf{Z})\right)$
around $\mathbb{E}_{\mathbf{Z}}\ft 1(\mathbf{Z})$ and $\ez\ft 2(\mathbf{Z})$
gives
\begin{eqnarray}
g\left(\ft 1(\mathbf{Z}),\ft 2(\mathbf{Z})\right) & = & \sum_{i=0}^{\lambda}\sum_{j=0}^{\lambda}\left(\left.\frac{\partial^{i+j}g(x,y)}{\partial x^{i}\partial y^{j}}\right|_{\substack{x=\ez\ft 1(\mathbf{Z})\\
y=\ez\ft 2(\mathbf{Z})
}
}\right)\frac{\et 1^{i}(\mathbf{Z})\et 2^{j}(\mathbf{Z})}{i!j!}+o\left(\et 1^{\lambda}(\mathbf{Z})+\et 2^{\lambda}(\mathbf{Z})\right)\label{eq:g_taylor2}
\end{eqnarray}
where $\et i(\mathbf{Z}):=\ft i(\mathbf{Z})-\mathbb{E}_{\mathbf{Z}}\ft i(\mathbf{Z})$.
To control the variance term in (\ref{eq:gsplit}), we thus require
expressions for $\ez\left[\et i^{j}(\mathbf{Z})\right]$. 

\begin{lemma}\label{lem:ekhat} \emph{Let $\mathbf{Z}$ be a realization
of the density $f_{2}$ that is in the interior of the support and
is independent of $\ft i$ for $i=1,2$. Let $n(q)$ be the set of
integer divisors of q including $1$ but excluding $q$. Then,
\begin{eqnarray}
\ez\left[\et i^{q}(\mathbf{Z})\right] & = & \begin{cases}
\sum_{j\in n(q)}\frac{1}{\left(N_{2}h_{2}^{d}\right)^{q-j}}\sum_{m=0}^{\left\lfloor s/2\right\rfloor }c_{6,i,q,j,m}(\mathbf{Z})h_{i}^{2m}+O\left(\frac{1}{N_{i}}\right), & q\geq2\\
0, & q=1,
\end{cases}\label{eq:moment}\\
\ez\left[\et 1^{q}(\mathbf{Z})\et 2^{l}(\mathbf{Z})\right] & = & \begin{cases}
\left(\sum_{i\in n(q)}\frac{1}{\left(N_{1}h_{1}^{d}\right)^{q-i}}\sum_{m=0}^{\left\lfloor s/2\right\rfloor }c_{6,1,q,i,m}(\mathbf{Z})h_{1}^{2m}\right)\times & q,\,l\geq2\\
\left(\sum_{j\in n(l)}\frac{1}{\left(N_{2}h_{2}^{d}\right)^{l-j}}\sum_{t=0}^{\left\lfloor s/2\right\rfloor }c_{6,2,l,j,t}(\mathbf{Z})h_{2}^{2t}\right)+O\left(\frac{1}{N_{1}}+\frac{1}{N_{2}}\right),\\
0, & q=1\,\text{or }l=1
\end{cases}\label{eq:cross_moment}
\end{eqnarray}
where $c_{6,i,q,j,m}$ is a functional of $f_{1}$ and $f_{2}.$}\end{lemma}
\begin{IEEEproof}
Define the random variable $\mathbf{V}_{i}(\mathbf{Z})=K\left(\frac{\mathbf{X}_{i}-\mathbf{Z}}{h_{2}}\right)-\ez K\left(\frac{\mathbf{X}_{i}-\mathbf{Z}}{h_{2}}\right)$.
This gives 
\begin{eqnarray*}
\et 2(\mathbf{Z}) & = & \ft 2(\mathbf{Z})-\ez\ft 2(\mathbf{Z})\\
 & = & \frac{1}{N_{2}h_{2}^{d}}\sum_{i=1}^{N_{2}}\mathbf{V}_{i}(\mathbf{Z}).
\end{eqnarray*}
Clearly, $\ez\mathbf{V}_{i}(\mathbf{Z})=0$. From (\ref{eq:substitution}),
we have for integer $j\geq1$ 
\begin{eqnarray*}
\ez\left[K^{j}\left(\frac{\mathbf{X}_{i}-\mathbf{Z}}{h_{2}}\right)\right] & = & \int K^{j}\left(t\right)f_{2}(th_{2}+\mathbf{Z})dt\\
 & = & h_{2}^{d}\sum_{m=0}^{\left\lfloor s/2\right\rfloor }c_{3,2,j,m}(\mathbf{Z})h_{2}^{2m},
\end{eqnarray*}
where the constants $c_{3,2,j,m}$ depend on the density $f_{2}$,
its derivatives, and the moments of the kernel $K^{j}$. Note that
since $K$ is symmetric, the odd moments of $K^{j}$ are zero for
$\mathbf{Z}$ in the interior of the support. However, all even moments
may now be nonzero since $K^{j}$ may now be nonnegative. By the binomial
theorem, 
\begin{eqnarray*}
\ez\left[\mathbf{V}_{i}^{j}(\mathbf{Z})\right] & = & \sum_{k=0}^{j}\binom{j}{k}\ez\left[K^{k}\left(\frac{\mathbf{X}_{i}-\mathbf{Z}}{h_{2}}\right)\right]\ez\left[K\left(\frac{\mathbf{X}_{i}-\mathbf{Z}}{h_{2}}\right)\right]^{j-k}\\
 & = & \sum_{k=0}^{j}\binom{j}{k}h_{2}^{d}O\left(h_{2}^{d(j-k)}\right)\sum_{m=0}^{\left\lfloor s/2\right\rfloor }c_{3,2,k,m}(\mathbf{Z})h_{2}^{2m}\\
 & = & h_{2}^{d}\sum_{m=0}^{\left\lfloor s/2\right\rfloor }c_{3,2,j,m}(\mathbf{Z})h_{2}^{2m}+O\left(h^{2d}\right).
\end{eqnarray*}
We can use these expressions to simplify $\ez\left[\et 2^{q}(\mathbf{Z})\right]$.
As an example, let $q=2$. Then since the $\mathbf{X}_{i}s$ are independent,
\begin{eqnarray*}
\ez\left[\et 2^{2}(\mathbf{Z})\right] & = & \frac{1}{N_{2}h_{2}^{2d}}\ez\mathbf{V}_{i}^{2}(\mathbf{Z})\\
 & = & \frac{1}{N_{2}h_{2}^{d}}\sum_{m=0}^{\left\lfloor s/2\right\rfloor }c_{3,2,2,m}(\mathbf{Z})h_{2}^{2m}+O\left(\frac{1}{N_{2}}\right).
\end{eqnarray*}
Similarly, we find that 
\begin{eqnarray*}
\ez\left[\et 2^{3}(\mathbf{Z})\right] & = & \frac{1}{N_{2}^{2}h_{2}^{3d}}\ez\mathbf{V}_{i}^{3}(\mathbf{Z})\\
 & = & \frac{1}{\left(N_{2}h_{2}^{d}\right)^{2}}\sum_{m=0}^{\left\lfloor s/2\right\rfloor }c_{3,2,3,m}(\mathbf{Z})h_{2}^{2m}+o\left(\frac{1}{N_{2}}\right).
\end{eqnarray*}
For $q=4$, we have 
\begin{eqnarray*}
\ez\left[\et 2^{4}(\mathbf{Z})\right] & = & \frac{1}{N_{2}^{3}h_{2}^{4d}}\ez\mathbf{V}_{i}^{4}(\mathbf{Z})+\frac{N_{2}-1}{N_{2}^{3}h_{2}^{4d}}\left(\ez\mathbf{V}_{i}^{2}(\mathbf{Z})\right)^{2}\\
 & = & \frac{1}{\left(N_{2}h_{2}^{d}\right)^{3}}\sum_{m=0}^{\left\lfloor s/2\right\rfloor }c_{3,2,4,m}(\mathbf{Z})h_{2}^{2m}+\frac{1}{\left(N_{2}h_{2}^{d}\right)^{2}}\sum_{m=0}^{\left\lfloor s/2\right\rfloor }c_{6,2,2,m}(\mathbf{Z})h_{2}^{2m}+o\left(\frac{1}{N_{2}}\right).
\end{eqnarray*}
The pattern is then for $q\geq2$, 
\[
\ez\left[\et 2^{q}(\mathbf{Z})\right]=\sum_{i\in n(q)}\frac{1}{\left(N_{2}h_{2}^{d}\right)^{q-i}}\sum_{m=0}^{\left\lfloor s/2\right\rfloor }c_{6,2,q,i,m}(\mathbf{Z})h_{2}^{2m}+O\left(\frac{1}{N_{2}}\right).
\]
 For any integer $q$, the largest possible factor is $q/2$. Thus
for given $q$, the smallest possible exponent on the $N_{2}h_{2}^{d}$
term is $q/2$. This increases as $q$ increases. A similar expression
holds for $\ez\left[\et 1^{q}(\mathbf{Z})\right]$ except the $\mathbf{X}_{i}$s
are replaced with $\mathbf{Y}_{i}$, $f_{2}$ is replaced with $f_{1}$,
and $N_{2}$ and $h_{2}$ are replaced with $N_{1}$ and $h_{1}$,
respectively, all resulting in different constants. Then since $\et 1(\mathbf{Z})$
and $\et 2(\mathbf{Z})$ are conditionally independent given $\mathbf{Z}$,
\begin{eqnarray*}
\ez\left[\et 1^{q}(\mathbf{Z})\et 2^{l}(\mathbf{Z})\right] & = & \left(\sum_{i\in n(q)}\frac{1}{\left(N_{1}h_{1}^{d}\right)^{q-i}}\sum_{m=0}^{\left\lfloor s/2\right\rfloor }c_{6,1,q,i,m}(\mathbf{Z})h_{1}^{2m}\right)\left(\sum_{j\in n(l)}\frac{1}{\left(N_{2}h_{2}^{d}\right)^{l-j}}\sum_{t=0}^{\left\lfloor s/2\right\rfloor }c_{6,2,l,j,t}(\mathbf{Z})h_{2}^{2t}\right)\\
 &  & +O\left(\frac{1}{N_{1}}+\frac{1}{N_{2}}\right).
\end{eqnarray*}
\end{IEEEproof}
Applying Lemma~\ref{lem:ekhat} to (\ref{eq:g_taylor2}) when taking
the conditional expectation given $\mathbf{Z}$ in the interior gives
an expression of the form
\begin{eqnarray}
\sum_{j=1}^{\lambda/2}\sum_{m=0}^{\left\lfloor s/2\right\rfloor }\left(c_{7,1,j,m}\left(\ez\ft 1(\mathbf{Z}),\ez\ft 2(\mathbf{Z})\right)\frac{h_{1}^{2m}}{\left(N_{1}h_{1}^{d}\right)^{j}}+c_{7,2,j,m}\left(\ez\ft 2(\mathbf{Z}),\ez\ft 2(\mathbf{Z})\right)\frac{h_{2}^{2m}}{\left(N_{2}h_{2}^{d}\right)^{j}}\right)\nonumber \\
+\sum_{j=1}^{\lambda/2}\sum_{m=0}^{\left\lfloor s/2\right\rfloor }\sum_{i=1}^{\lambda/2}\sum_{n=0}^{\left\lfloor s/2\right\rfloor }c_{7,j,i,m,n}\left(\ez\ft 2(\mathbf{Z}),\ez\ft 2(\mathbf{Z})\right)\frac{h_{1}^{2m}h_{2}^{2n}}{\left(N_{1}h_{1}^{d}\right)^{j}\left(N_{2}h_{2}^{d}\right)^{i}}\nonumber \\
+O\left(\frac{1}{\left(N_{1}h_{1}^{d}\right)^{\frac{\lambda}{2}}}+\frac{1}{\left(N_{2}h_{2}^{d}\right)^{\frac{\lambda}{2}}}\right).\label{eq:interior_ez}
\end{eqnarray}

Note that the functionals $c_{7,i,j,m}$ and $c_{7,j,i,m,n}$ depend
on the derivatives of $g$ and $\ez\ft i(\mathbf{Z})$ which depends
on $h_{i}$. To apply ensemble estimation, we need to separate the
dependence on $h_{i}$ from the constants. If we use ODin1, then it
is sufficient to note that in the interior of the support, $\ez\ft i(\mathbf{Z})=f_{i}(\mathbf{Z})+o(1)$
and therefore $c\left(\ez\ft 1(\mathbf{Z}),\ez\ft 2(\mathbf{Z})\right)=c\left(f_{1}(\mathbf{Z}),f_{2}(\mathbf{Z})\right)+o(1)$
for some functional $c$. The terms in (\ref{eq:interior_ez}) reduce
to 
\[
c_{7,1,1,0}\left(f_{1}(\mathbf{Z}),f_{2}(\mathbf{Z})\right)\frac{1}{N_{1}h_{1}^{d}}+c_{7,2,1,0}\left(f_{1}(\mathbf{Z}),f_{2}(\mathbf{Z})\right)\frac{1}{N_{2}h_{2}^{d}}+o\left(\frac{1}{N_{1}h_{1}^{d}}+\frac{1}{N_{2}h_{2}^{d}}\right).
\]
For ODin2, we need the higher order terms. To separate the dependence
on $h_{i}$ from the constants, we need more information about the
functional $g$ and its derivatives. Consider the special case where
the functional $g(x,y)$ has derivatives of the form of $x^{\alpha}y^{\beta}$
with $\alpha,\beta<0.$ This includes the important cases of the KL
divergence and the Renyi divergence. The generalized binomial theorem
states that if $\binom{\alpha}{m}:=\frac{\alpha(\alpha-1)\dots(\alpha-m+1)}{m!}$
and if $q$ and $t$ are real numbers with $|q|>|t|$, then for any
complex number $\alpha$, 
\begin{equation}
(q+t)^{\alpha}=\sum_{m=0}^{\infty}\binom{\alpha}{m}q^{\alpha-m}t^{m}.\label{eq:general_binomial}
\end{equation}
Since the densities are bounded away from zero, for sufficiently small
$h_{i}$, we have that $f_{i}(\mathbf{Z})>\left|\sum_{j=\nu/2}^{\left\lfloor s/2\right\rfloor }c_{i,j}(\mathbf{Z})h_{i}^{2j}+O\left(h_{i}^{s}\right)\right|.$
Applying the generalized binomial theorem and Lemma~\ref{lem:interior}
gives that 
\[
\left(\ez\ft 1(\mathbf{Z})\right)^{\alpha}=\sum_{m=0}^{\infty}\binom{\alpha}{m}f_{i}^{\alpha-m}(\mathbf{Z})\left(\sum_{j=\nu/2}^{\left\lfloor s/2\right\rfloor }c_{i,j}(\mathbf{Z})h_{i}^{2j}+O\left(h_{i}^{s}\right)\right)^{m}.
\]
Since $m$ is an integer, the exponents of the $h_{i}$ terms are
also integers. Thus (\ref{eq:interior_ez}) gives in this case 
\begin{eqnarray}
\ez\left[g\left(\ft 1(\mathbf{Z}),\ft 2(\mathbf{Z})\right)-g\left(\ez\ft 1(\mathbf{Z}),\ez\ft 2(\mathbf{Z})\right)\right] & = & \sum_{j=1}^{\lambda/2}\sum_{m=0}^{\left\lfloor s/2\right\rfloor }\left(c_{8,1,j,m}\left(\mathbf{Z}\right)\frac{h_{1}^{2m}}{\left(N_{1}h_{1}^{d}\right)^{j}}+c_{8,2,j,m}\left(\mathbf{Z}\right)\frac{h_{2}^{2m}}{\left(N_{2}h_{2}^{d}\right)^{j}}\right)\nonumber \\
 &  & +\sum_{j=1}^{\lambda/2}\sum_{m=0}^{\left\lfloor s/2\right\rfloor }\sum_{i=1}^{\lambda/2}\sum_{n=0}^{\left\lfloor s/2\right\rfloor }c_{8,j,i,m,n}\left(\mathbf{Z}\right)\frac{h_{1}^{2m}h_{2}^{2n}}{\left(N_{1}h_{1}^{d}\right)^{j}\left(N_{2}h_{2}^{d}\right)^{i}}\nonumber \\
 &  & +O\left(\frac{1}{\left(N_{1}h_{1}^{d}\right)^{\frac{\lambda}{2}}}+\frac{1}{\left(N_{2}h_{2}^{d}\right)^{\frac{\lambda}{2}}}+h_{1}^{s}+h_{2}^{s}\right).\label{eq:interior_ez2}
\end{eqnarray}

As before, the case for $\mathbf{Z}$ close to the boundary of the
support is more complicated. However, by using a similar technique
to the proof of Lemma~\ref{lem:boundary} for $\mathbf{Z}$ at the
boundary and combining with the previous results, we find that for
general $g$, 
\begin{equation}
\bE\left[g\left(\ft 1(\mathbf{Z}),\ft 2(\mathbf{Z})\right)-g\left(\ez\ft 1(\mathbf{Z}),\ez\ft 2(\mathbf{Z})\right)\right]=c_{9,1}\frac{1}{N_{1}h_{1}^{d}}+c_{9,2}\frac{1}{N_{2}h_{2}^{d}}+o\left(\frac{1}{N_{1}h_{1}^{d}}+\frac{1}{N_{2}h_{2}^{d}}\right).\label{eq:bias_result2}
\end{equation}
If $g(x,y)$ has derivatives of the form of $x^{\alpha}y^{\beta}$
with $\alpha,\beta<0$, then we can similarly obtain 
\begin{eqnarray}
\bE\left[g\left(\ft 1(\mathbf{Z}),\ft 2(\mathbf{Z})\right)-g\left(\ez\ft 1(\mathbf{Z}),\ez\ft 2(\mathbf{Z})\right)\right] & = & \sum_{j=1}^{\lambda/2}\sum_{m=0}^{r}\left(c_{9,1,j,m}\frac{h_{1}^{m}}{\left(N_{1}h_{1}^{d}\right)^{j}}+c_{9,2,j,m}\frac{h_{2}^{m}}{\left(N_{2}h_{2}^{d}\right)^{j}}\right)\nonumber \\
 &  & +\sum_{j=1}^{\lambda/2}\sum_{m=0}^{r}\sum_{i=1}^{\lambda/2}\sum_{n=0}^{r}c_{9,j,i,m,n}\frac{h_{1}^{m}h_{2}^{n}}{\left(N_{1}h_{1}^{d}\right)^{j}\left(N_{2}h_{2}^{d}\right)^{i}}\nonumber \\
 &  & +O\left(\frac{1}{\left(N_{1}h_{1}^{d}\right)^{\frac{\lambda}{2}}}+\frac{1}{\left(N_{2}h_{2}^{d}\right)^{\frac{\lambda}{2}}}+h_{1}^{s}+h_{2}^{s}\right).\label{eq:bias_result3}
\end{eqnarray}
Combining (\ref{eq:biasresult1}) with either (\ref{eq:bias_result2})
or (\ref{eq:bias_result3}) completes the proof.

\section{Proof of Theorem~\ref{thm:var-1} (Variance)}

\label{sec:VarProof}To bound the variance of the plug-in estimator
$\gt$, we will use the Efron-Stein inequality~\cite{efron1981jackknife}:

\begin{lemma}[Efron-Stein Inequality] \emph{Let $\mathbf{X}_{1},\dots,\mathbf{X}_{n},\mathbf{X}_{1}^{'},\dots,\mathbf{X}_{n}^{'}$
be independent random variables on the space $\mathcal{S}$. Then
if $f:\mathcal{S}\times\dots\times\mathcal{S}\rightarrow\mathbb{R}$,
we have that 
\[
\var\left[f(\mathbf{X}_{1},\dots,\mathbf{X}_{n})\right]\leq\frac{1}{2}\sum_{i=1}^{n}\bE\left[\left(f(\mathbf{X}_{1},\dots,\mathbf{X}_{n})-f(\mathbf{X}_{1},\dots,\mathbf{X}_{i}^{'},\dots,\mathbf{X}_{n})\right)^{2}\right].
\]
}

\end{lemma}

Suppose we have samples $\left\{ \mathbf{X}_{1},\dots,\mathbf{X}_{N_{2}},\mathbf{Y}_{1},\dots,\mathbf{Y}_{N_{1}}\right\} $
and $\left\{ \mathbf{X}_{1}^{'},\dots,\mathbf{X}_{N_{2}},\mathbf{Y}_{1},\dots,\mathbf{Y}_{N_{1}}\right\} $
and denote the respective estimators as $\gt$ and $\gt^{'}$. We
have that 
\begin{eqnarray}
\left|\gt-\gt^{'}\right| & \leq & \frac{1}{N_{2}}\left|g\left(\ft 1(\mathbf{X}_{1}),\ft 2(\mathbf{X}_{1})\right)-g\left(\ft 1(\mathbf{X}_{1}^{'}),\ft 2(\mathbf{X}_{1}^{'})\right)\right|\nonumber \\
 &  & +\frac{1}{N_{2}}\sum_{j=2}^{N_{2}}\left|g\left(\ft 1(\mathbf{X}_{j}),\ft 2(\mathbf{X}_{j})\right)-g\left(\ft 1(\mathbf{X}_{j}),\ft 2^{'}(\mathbf{X}_{j})\right)\right|.\label{eq:triangle}
\end{eqnarray}
Since $g$ is Lipschitz continuous with constant $C_{g}$, we have
\begin{eqnarray}
\left|g\left(\ft 1(\mathbf{X}_{1}),\ft 2(\mathbf{X}_{1})\right)-g\left(\ft 1(\mathbf{X}_{1}^{'}),\ft 2(\mathbf{X}_{1}^{'})\right)\right| & \leq & C_{g}\left(\left|\ft 1(\mathbf{X}_{1})-\ft 1(\mathbf{X}_{1}^{'})\right|+\left|\ft 2(\mathbf{X}_{1})-\ft 2(\mathbf{X}_{1}^{'})\right|\right),\label{eq:lipschitz}
\end{eqnarray}
\begin{eqnarray}
\left|\ft 1(\mathbf{X}_{1})-\ft 1(\mathbf{X}_{1}^{'})\right| & = & \frac{1}{N_{1}h_{1}^{d}}\left|\sum_{i=1}^{N_{1}}\left(K\left(\frac{\mathbf{X}_{1}-\mathbf{Y}_{i}}{h_{1}}\right)-K\left(\frac{\mathbf{X}_{1}^{'}-\mathbf{Y}_{i}}{h_{1}}\right)\right)\right|\nonumber \\
 & \leq & \frac{1}{N_{1}h_{1}^{d}}\sum_{i=1}^{N_{1}}\left|K\left(\frac{\mathbf{X}_{1}-\mathbf{Y}_{i}}{h_{1}}\right)-K\left(\frac{\mathbf{X}_{1}^{'}-\mathbf{Y}_{i}}{h_{1}}\right)\right|\nonumber \\
\implies\bE\left[\left|\ft 1(\mathbf{X}_{1})-\ft 1(\mathbf{X}_{1}^{'})\right|^{2}\right] & \leq & \frac{1}{N_{1}h_{1}^{2d}}\sum_{i=1}^{N_{1}}\bE\left[\left(K\left(\frac{\mathbf{X}_{1}-\mathbf{Y}_{i}}{h_{1}}\right)-K\left(\frac{\mathbf{X}_{1}^{'}-\mathbf{Y}_{i}}{h_{1}}\right)\right)^{2}\right],\label{eq:densityBound}
\end{eqnarray}
where the last step follows from Jensen's inequality. By making the
substitution $\mathbf{u}_{i}=\frac{\mathbf{X}_{1}-\mathbf{Y}_{i}}{h_{1}}$
and $\mathbf{u}_{i}^{'}=\frac{\mathbf{X}_{1}^{'}-\mathbf{Y}_{i}}{h_{1}}$,
this gives 
\begin{eqnarray*}
\frac{1}{h_{1}^{2d}}\bE\left[\left(K\left(\frac{\mathbf{X}_{1}-\mathbf{Y}_{i}}{h_{1}}\right)-K\left(\frac{\mathbf{X}_{1}^{'}-\mathbf{Y}_{i}}{h_{1}}\right)\right)^{2}\right] & = & \frac{1}{h^{2d}}\int\left(K\left(\frac{\mathbf{X}_{1}-\mathbf{Y}_{i}}{h_{1}}\right)-K\left(\frac{\mathbf{X}_{1}^{'}-\mathbf{Y}_{i}}{h_{1}}\right)\right)^{2}\times\\
 &  & f_{2}(\mathbf{X}_{1})f_{2}(\mathbf{X}_{1}^{'})f_{1}(\mathbf{Y}_{i})d\mathbf{X}_{1}d\mathbf{X}_{1}^{'}d\mathbf{Y}_{i}\\
 & \leq & 2||K||_{\infty}^{2}.
\end{eqnarray*}
Combining this with (\ref{eq:densityBound}) gives 
\[
\bE\left[\left|\ft 1(\mathbf{X}_{1})-\ft 1(\mathbf{X}_{1}^{'})\right|^{2}\right]\leq2||K||_{\infty}^{2}.
\]
Similarly, 
\[
\bE\left[\left|\ft 2(\mathbf{X}_{1})-\ft 2(\mathbf{X}_{1}^{'})\right|^{2}\right]\leq2||K||_{\infty}^{2}.
\]
Combining these results with (\ref{eq:lipschitz}) gives 
\begin{equation}
\bE\left[\left(g\left(\ft 1(\mathbf{X}_{1}),\ft 2(\mathbf{X}_{1})\right)-g\left(\ft 1(\mathbf{X}_{1}^{'}),\ft 2(\mathbf{X}_{1}^{'})\right)\right)^{2}\right]\leq8C_{g}^{2}||K||_{\infty}^{2}.\label{eq:1stTermFinal}
\end{equation}

The second term in (\ref{eq:triangle}) is controlled in a similar
way. From the Lipschitz condition, 
\begin{eqnarray*}
\left|g\left(\ft 1(\mathbf{X}_{j}),\ft 2(\mathbf{X}_{j})\right)-g\left(\ft 1(\mathbf{X}_{j}),\ft 2^{'}(\mathbf{X}_{j})\right)\right|^{2} & \leq & C_{g}^{2}\left|\ft 2(\mathbf{X}_{j})-\ft 2^{'}(\mathbf{X}_{j})\right|^{2}\\
 & = & \frac{C_{g}^{2}}{M_{2}^{2}h_{2}^{2d}}\left(K\left(\frac{\mathbf{X}_{j}-\mathbf{X}_{1}}{h}\right)-K\left(\frac{\mathbf{X}_{j}-\mathbf{X}_{1}^{'}}{h}\right)\right)^{2}.
\end{eqnarray*}
The $h_{2}^{2d}$ terms are eliminated by making the substitutions
of $\mathbf{u}_{j}=\frac{\mathbf{X}_{j}-\mathbf{X}_{1}}{h_{2}}$ and
$\mathbf{u}_{j}^{'}=\frac{\mathbf{X}_{j}-\mathbf{X}_{1}^{'}}{h_{2}}$
within the expectation to obtain 
\begin{eqnarray}
\bE\left[\left|g\left(\ft 1(\mathbf{X}_{j}),\ft 2(\mathbf{X}_{j})\right)-g\left(\ft 1(\mathbf{X}_{j}),\ft 2^{'}(\mathbf{X}_{j})\right)\right|^{2}\right] & \leq & \frac{2C_{g}^{2}||K||_{\infty}^{2}}{M_{2}^{2}}\label{eq:gsquared}
\end{eqnarray}
\[
\implies\bE\left[\left(\sum_{j=2}^{N_{2}}\left|g\left(\ft 1(\mathbf{X}_{j}),\ft 2(\mathbf{X}_{j})\right)-g\left(\ft 1(\mathbf{X}_{j}),\ft 2^{'}(\mathbf{X}_{j})\right)\right|\right)^{2}\right]
\]
\begin{eqnarray}
 & = & \sum_{j=2}^{N_{2}}\sum_{i=2}^{N_{2}}\bE\left[\left|g\left(\ft 1(\mathbf{X}_{j}),\ft 2(\mathbf{X}_{j})\right)-g\left(\ft 1(\mathbf{X}_{j}),\ft 2^{'}(\mathbf{X}_{j})\right)\right|\left|g\left(\ft 1(\mathbf{X}_{i}),\ft 2(\mathbf{X}_{i})\right)-g\left(\ft 1(\mathbf{X}_{i}),\ft 2^{'}(\mathbf{X}_{i})\right)\right|\right]\nonumber \\
 & \leq & M_{2}^{2}\bE\left[\left|g\left(\ft 1(\mathbf{X}_{j}),\ft 2(\mathbf{X}_{j})\right)-g\left(\ft 1(\mathbf{X}_{j}),\ft 2^{'}(\mathbf{X}_{j})\right)\right|^{2}\right]\nonumber \\
 & \leq & 2C_{g}^{2}||K||_{\infty}^{2},\label{eq:2ndtermFinal}
\end{eqnarray}
where we use the Cauchy Schwarz inequality to bound the expectation
within each summand. Finally, applying Jensen's inequality and (\ref{eq:1stTermFinal})
and (\ref{eq:2ndtermFinal}) gives 
\begin{eqnarray*}
\bE\left[\left|\gt-\gt^{'}\right|^{2}\right] & \leq & \frac{2}{N_{2}^{2}}\bE\left[\left|g\left(\ft 1(\mathbf{X}_{1}),\ft 2(\mathbf{X}_{1})\right)-g\left(\ft 1(\mathbf{X}_{1}^{'}),\ft 2(\mathbf{X}_{1}^{'})\right)\right|^{2}\right]\\
 &  & +\frac{2}{N_{2}^{2}}\bE\left[\left(\sum_{j=2}^{N_{2}}\left|g\left(\ft 1(\mathbf{X}_{j}),\ft 2(\mathbf{X}_{j})\right)-g\left(\ft 1(\mathbf{X}_{j}),\ft 2^{'}(\mathbf{X}_{j})\right)\right|\right)^{2}\right]\\
 & \leq & \frac{20C_{g}^{2}||K||_{\infty}^{2}}{N_{2}^{2}}.
\end{eqnarray*}

Now suppose we have samples $\left\{ \mathbf{X}_{1},\dots,\mathbf{X}_{N_{2}},\mathbf{Y}_{1},\dots,\mathbf{Y}_{N_{1}}\right\} $
and $\left\{ \mathbf{X}_{1},\dots,\mathbf{X}_{N_{2}},\mathbf{Y}_{1}^{'},\dots,\mathbf{Y}_{N_{1}}\right\} $
and denote the respective estimators as $\gt$ and $\gt^{'}$. Then
\begin{eqnarray*}
\left|g\left(\ft 1(\mathbf{X}_{j}),\ft 2(\mathbf{X}_{j})\right)-g\left(\ft 1^{'}(\mathbf{X}_{j}),\ft 2(\mathbf{X}_{j})\right)\right| & \leq & C_{g}\left|\ft 1(\mathbf{X}_{j})-\ft 1^{'}(\mathbf{X}_{j})\right|\\
 & = & \frac{C_{g}}{N_{1}h_{1}^{d}}\left|K\left(\frac{\mathbf{X}_{j}-\mathbf{Y}_{1}}{h_{1}}\right)-K\left(\frac{\mathbf{X}_{j}-\mathbf{Y}_{1}^{'}}{h_{1}}\right)\right|\\
\implies\bE\left[\left|g\left(\ft 1(\mathbf{X}_{j}),\ft 2(\mathbf{X}_{j})\right)-g\left(\ft 1^{'}(\mathbf{X}_{j}),\ft 2(\mathbf{X}_{j})\right)\right|^{2}\right] & \leq & \frac{2C_{g}^{2}||K||_{\infty}^{2}}{N_{1}^{2}}.
\end{eqnarray*}
Thus using a similar argument as was used to obtain (\ref{eq:2ndtermFinal}),
\begin{eqnarray*}
\bE\left[\left|\gt-\gt^{'}\right|^{2}\right] & \leq & \frac{1}{N_{2}^{2}}\bE\left[\left(\sum_{j=1}^{N_{2}}\left|g\left(\ft 1(\mathbf{X}_{j}),\ft 2(\mathbf{X}_{j})\right)-g\left(\ft 1^{'}(\mathbf{X}_{j}),\ft 2(\mathbf{X}_{j})\right)\right|\right)^{2}\right]\\
 & \leq & \frac{2C_{g}^{2}||K||_{\infty}^{2}}{N_{2}^{2}}.
\end{eqnarray*}
Applying the Efron-Stein inequality gives 
\[
\var\left[\gt\right]\leq\frac{10C_{g}^{2}||K||_{\infty}^{2}}{N_{2}}+\frac{C_{g}^{2}||K||_{\infty}^{2}N_{1}}{N_{2}^{2}}.
\]

\section{Proof of Theorem~\ref{thm:clt} (CLT)}

\label{sec:cltProof}We are interested in the asymptotic distribution
of 
\begin{eqnarray*}
\sqrt{N_{2}}\left(\gt-\bE\left[\gt\right]\right) & = & \frac{1}{\sqrt{N_{2}}}\sum_{j=1}^{N_{2}}\left(g\left(\ft 1(\mathbf{X}_{j}),\ft 2(\mathbf{X}_{j})\right)-\bE_{\mathbf{X}_{j}}\left[g\left(\ft 1(\mathbf{X}_{j}),\ft 2(\mathbf{X}_{j})\right)\right]\right)\\
 &  & +\frac{1}{\sqrt{N_{2}}}\sum_{j=1}^{N_{2}}\left(\bE_{\mathbf{X}_{j}}\left[g\left(\ft 1(\mathbf{X}_{j}),\ft 2(\mathbf{X}_{j})\right)\right]-\bE\left[g\left(\ft 1(\mathbf{X}_{j}),\ft 2(\mathbf{X}_{j})\right)\right]\right).
\end{eqnarray*}
Note that by the standard central limit theorem~\cite{durrett2010probability},
the second term converges in distribution to a Gaussian random variable.
If the first term converges in probability to a constant (specifically,
0), then we can use Slutsky's theorem~\cite{gut2012probability}\textbf{
}to find the asymptotic distribution. So now we focus on the first
term which we denote as $\mathbf{V}_{N_{2}}$.

To prove convergence in probability, we will use Chebyshev's inequality.
Note that $\bE\left[\mathbf{V}_{N_{2}}\right]=0$. To bound the variance
of $\mathbf{V}_{N_{2}}$, we again use the Efron-Stein inequality.
Let $\mathbf{X}_{1}^{'}$ be drawn from $f_{2}$ and denote $\mathbf{V}_{N_{2}}$
and $\mathbf{V}_{N_{2}}^{'}$ as the sequences using $\mathbf{X}_{1}$
and $\mathbf{X}_{1}^{'}$, respectively. Then 
\begin{eqnarray}
\mathbf{V}_{N_{2}}-\mathbf{V}_{N_{2}}^{'} & = & \frac{1}{\sqrt{N_{2}}}\left(g\left(\ft 1(\mathbf{X}_{1}),\ft 2(\mathbf{X}_{1})\right)-\bE_{\mathbf{X}_{1}}\left[g\left(\ft 1(\mathbf{X}_{1}),\ft 2(\mathbf{X}_{1})\right)\right]\right)\nonumber \\
 &  & -\frac{1}{\sqrt{N_{2}}}\left(g\left(\ft 1(\mathbf{X}_{1}^{'}),\ft 2(\mathbf{X}_{1}^{'})\right)-\bE_{\mathbf{X}_{1}^{'}}\left[g\left(\ft 1(\mathbf{X}_{1}^{'}),\ft 2(\mathbf{X}_{1}^{'})\right)\right]\right)\nonumber \\
 &  & +\frac{1}{\sqrt{N_{2}}}\sum_{j=2}^{N_{2}}\left(g\left(\ft 1(\mathbf{X}_{j}),\ft 2(\mathbf{X}_{j})\right)-g\left(\ft 1(\mathbf{X}_{j}),\ft 2^{'}(\mathbf{X}_{j})\right)\right).\label{eq:VnDiff}
\end{eqnarray}
Note that 
\[
\bE\left[\left(g\left(\ft 1(\mathbf{X}_{1}),\ft 2(\mathbf{X}_{1})\right)-\bE_{\mathbf{X}_{1}}\left[g\left(\ft 1(\mathbf{X}_{1}),\ft 2(\mathbf{X}_{1})\right)\right]\right)^{2}\right]=\bE\left[\var_{\mathbf{X}_{1}}\left[g\left(\ft 1(\mathbf{X}_{1}),\ft 2(\mathbf{X}_{1})\right)\right]\right].
\]
We will use the Efron-Stein inequality to bound $\var_{\mathbf{X}_{1}}\left[g\left(\ft 1(\mathbf{X}_{1}),\ft 2(\mathbf{X}_{1})\right)\right]$.
We do this by bounding the conditional expectation of the term 
\[
\left|g\left(\ft 1(\mathbf{X}_{1}),\ft 2(\mathbf{X}_{1})\right)-g\left(\ft 1(\mathbf{X}_{1}),\ft 2^{'}(\mathbf{X}_{1})\right)\right|,
\]
where $\mathbf{X}_{i}$ is replaced with $\mathbf{X}_{i}^{'}$ in
the KDEs for some $i\neq1$. Using similar steps as in Appendix~\ref{sec:VarProof},
we have that 
\[
\bE_{\mathbf{X}_{1}}\left[\left|g\left(\ft 1(\mathbf{X}_{1}),\ft 2(\mathbf{X}_{1})\right)-g\left(\ft 1(\mathbf{X}_{1}),\ft 2^{'}(\mathbf{X}_{1})\right)\right|^{2}\right]=O\left(\frac{1}{N_{2}^{2}}\right).
\]
A similar result is obtained when $\mathbf{Y}_{i}$ is replaced with
$\mathbf{\mathbf{Y}_{i}^{'}}$. Then by the Efron-Stein inequality,
$\var_{\mathbf{X}_{1}}\left[g\left(\ft 1(\mathbf{X}_{1}),\ft 2(\mathbf{X}_{1})\right)\right]=O\left(\frac{1}{N_{2}}+\frac{1}{N_{1}}\right)$.
Therefore, 
\[
\bE\left[\frac{1}{N_{2}}\left(g\left(\ft 1(\mathbf{X}_{1}),\ft 2(\mathbf{X}_{1})\right)-\bE_{\mathbf{X}_{1}}\left[g\left(\ft 1(\mathbf{X}_{1}),\ft 2(\mathbf{X}_{1})\right)\right]\right)^{2}\right]=O\left(\frac{1}{N_{2}^{2}}+\frac{1}{N_{1}N_{2}}\right).
\]
A similar result holds for the $g\left(\ft 1(\mathbf{X}_{1}^{'}),\ft 2(\mathbf{X}_{1}^{'})\right)$
terms in (\ref{eq:VnDiff}).

For the third term in (\ref{eq:VnDiff}), 
\[
\bE\left[\left(\sum_{j=2}^{N_{2}}\left|g\left(\ft 1(\mathbf{X}_{j}),\ft 2(\mathbf{X}_{j})\right)-g\left(\ft 1(\mathbf{X}_{j}),\ft 2^{'}(\mathbf{X}_{j})\right)\right|\right)^{2}\right]
\]
\[
=\sum_{j=2}^{N_{2}}\sum_{i=2}^{N_{2}}\bE\left[\left|g\left(\ft 1(\mathbf{X}_{j}),\ft 2(\mathbf{X}_{j})\right)-g\left(\ft 1(\mathbf{X}_{j}),\ft 2^{'}(\mathbf{X}_{j})\right)\right|\left|g\left(\ft 1(\mathbf{X}_{i}),\ft 2(\mathbf{X}_{i})\right)-g\left(\ft 1(\mathbf{X}_{i}),\ft 2^{'}(\mathbf{X}_{i})\right)\right|\right].
\]
There are $M_{2}$ terms where $i=j$ and we have from Appendix~\ref{sec:VarProof}
(see (\ref{eq:gsquared})) that 
\[
\bE\left[\left|g\left(\ft 1(\mathbf{X}_{j}),\ft 2(\mathbf{X}_{j})\right)-g\left(\ft 1(\mathbf{X}_{j}),\ft 2^{'}(\mathbf{X}_{j})\right)\right|^{2}\right]\leq\frac{2C_{g}^{2}||K||_{\infty}^{2}}{M_{2}^{2}}.
\]
Thus these terms are $O\left(\frac{1}{M_{2}}\right)$. There are $M_{2}^{2}-M_{2}$
terms when $i\neq j$. In this case, we can do four substitutions
of the form $\mathbf{u}_{j}=\frac{\mathbf{X}_{j}-\mathbf{X}_{1}}{h_{2}}$
to obtain 
\[
\bE\left[\left|g\left(\ft 1(\mathbf{X}_{j}),\ft 2(\mathbf{X}_{j})\right)-g\left(\ft 1(\mathbf{X}_{j}),\ft 2^{'}(\mathbf{X}_{j})\right)\right|\left|g\left(\ft 1(\mathbf{X}_{i}),\ft 2(\mathbf{X}_{i})\right)-g\left(\ft 1(\mathbf{X}_{i}),\ft 2^{'}(\mathbf{X}_{i})\right)\right|\right]\leq\frac{4C_{g}^{2}||K||_{\infty}^{2}h_{2}^{2d}}{M_{2}^{2}}.
\]
Then since $h_{2}^{d}=o(1)$, we get 
\begin{equation}
\bE\left[\left(\sum_{j=2}^{N_{2}}\left|g\left(\ft 1(\mathbf{X}_{j}),\ft 2(\mathbf{X}_{j})\right)-g\left(\ft 1(\mathbf{X}_{j}),\ft 2^{'}(\mathbf{X}_{j})\right)\right|\right)^{2}\right]=o(1),\label{eq:CLTSum}
\end{equation}
\begin{eqnarray*}
\implies\bE\left[\left(\mathbf{V}_{N_{2}}-\mathbf{V}_{N_{2}}^{'}\right)^{2}\right] & \leq & \frac{3}{N_{2}}\bE\left[\left(g\left(\ft 1(\mathbf{X}_{1}),\ft 2(\mathbf{X}_{1})\right)-\bE_{\mathbf{X}_{1}}\left[g\left(\ft 1(\mathbf{X}_{1}),\ft 2(\mathbf{X}_{1})\right)\right]\right)^{2}\right]\\
 &  & +\frac{3}{N_{2}}\bE\left[\left(g\left(\ft 1(\mathbf{X}_{1}^{'}),\ft 2(\mathbf{X}_{1}^{'})\right)-\bE_{\mathbf{X}_{1}^{'}}\left[g\left(\ft 1(\mathbf{X}_{1}^{'}),\ft 2(\mathbf{X}_{1}^{'})\right)\right]\right)^{2}\right]\\
 &  & +\frac{3}{N_{2}}\bE\left[\left(\sum_{j=2}^{N_{2}}\left(g\left(\ft 1(\mathbf{X}_{j}),\ft 2(\mathbf{X}_{j})\right)-g\left(\ft 1^{'}(\mathbf{X}_{j}),\ft 2^{'}(\mathbf{X}_{j})\right)\right)\right)^{2}\right]\\
 & = & o\left(\frac{1}{N_{2}}\right).
\end{eqnarray*}

Now consider samples $\left\{ \mathbf{X}_{1},\dots,\mathbf{X}_{N_{2}},\mathbf{Y}_{1},\dots,\mathbf{Y}_{N_{1}}\right\} $
and $\left\{ \mathbf{X}_{1},\dots,\mathbf{X}_{N_{2}},\mathbf{Y}_{1}^{'},\dots,\mathbf{Y}_{N_{1}}\right\} $
and the respective sequences $\mathbf{V}_{N_{2}}$ and $\mathbf{V}_{N_{2}}^{'}$.
Then 
\begin{eqnarray*}
\mathbf{V}_{N_{2}}-\mathbf{V}_{N_{2}}^{'} & = & \frac{1}{\sqrt{N_{2}}}\sum_{j=1}^{N_{2}}\left(g\left(\ft 1(\mathbf{X}_{j}),\ft 2(\mathbf{X}_{j})\right)-g\left(\ft 1^{'}(\mathbf{X}_{j}),\ft 2(\mathbf{X}_{j})\right)\right).
\end{eqnarray*}
Using a similar argument as that used to obtain (\ref{eq:CLTSum}),
we have that if $h_{1}^{d}=o(1)$ and $N_{1}\rightarrow\infty$, then
\[
\bE\left[\left(\sum_{j=2}^{N_{2}}\left|g\left(\ft 1(\mathbf{X}_{j}),\ft 2(\mathbf{X}_{j})\right)-g\left(\ft 1^{'}(\mathbf{X}_{j}),\ft 2(\mathbf{X}_{j})\right)\right|\right)^{2}\right]=o(1)
\]
\[
\implies\bE\left[\left(\mathbf{V}_{N_{2}}-\mathbf{V}_{N_{2}}^{'}\right)^{2}\right]=o\left(\frac{1}{N_{2}}\right).
\]
Applying the Efron-Stein inequality gives 
\[
\var\left[\mathbf{V}_{N_{2}}\right]=o\left(\frac{N_{2}+N_{1}}{N_{2}}\right)=o(1).
\]
Thus by Chebyshev's inequality, 
\[
\Pr\left(\left|\mathbf{V}_{N_{2}}\right|>\epsilon\right)\leq\frac{\var\left[\mathbf{V}_{N_{2}}\right]}{\epsilon^{2}}=o(1),
\]
and therefore $\mathbf{V}_{N_{2}}$ converges to zero in probability.
By Slutsky's theorem, $\sqrt{N_{2}}\left(\gt-\bE\left[\gt\right]\right)$
converges in distribution to a zero mean Gaussian random variable
with variance 
\[
\var\left[\bE_{\mathbf{X}}\left[g\left(\ft 1(\mathbf{X}),\ft 2(\mathbf{X})\right)\right]\right],
\]
 where $\mathbf{X}$ is drawn from $f_{2}$.

For the weighted ensemble estimator, we wish to know the asymptotic
distribution of $\sqrt{N_{2}}\left(\tilde{\mathbf{G}}_{w}-\bE\left[\tilde{\mathbf{G}}_{w}\right]\right)$
where $\tilde{\mathbf{G}}_{w}=\sum_{l\in\bar{l}}w(l)\tilde{\mathbf{G}}_{h(l)}$.
We have that 
\begin{eqnarray*}
\sqrt{N_{2}}\left(\tilde{\mathbf{G}}_{w}-\bE\left[\tilde{\mathbf{G}}_{w}\right]\right) & = & \frac{1}{\sqrt{N_{2}}}\sum_{j=1}^{N_{2}}\sum_{l\in\bar{l}}w(l)\left(g\left(\ftl 1(\mathbf{X}_{j}),\ftl 2(\mathbf{X}_{j})\right)-\bE_{\mathbf{X}_{j}}\left[g\left(\ftl 1(\mathbf{X}_{j}),\ftl 2(\mathbf{X}_{j})\right)\right]\right)\\
 &  & +\frac{1}{\sqrt{N_{2}}}\sum_{j=1}^{N_{2}}\left(\bE_{\mathbf{X}_{j}}\left[\sum_{l\in\bar{l}}w(l)g\left(\ftl 1(\mathbf{X}_{j}),\ftl 2(\mathbf{X}_{j})\right)\right]-\bE\left[\sum_{l\in\bar{l}}w(l)g\left(\ftl 1(\mathbf{X}_{j}),\ftl 2(\mathbf{X}_{j})\right)\right]\right).
\end{eqnarray*}
The second term again converges in distribution to a Gaussian random
variable by the central limit theorem. The mean and variance are,
respectively, zero and 
\[
\var\left[\sum_{l\in\bar{l}}w(l)\bE_{\mathbf{X}}\left[g\left(\ftl 1(\mathbf{X}),\ftl 2(\mathbf{X})\right)\right]\right].
\]
The first term is equal to 
\begin{eqnarray*}
\sum_{l\in\bar{l}}w(l)\left(\frac{1}{\sqrt{N_{2}}}\sum_{j=1}^{N_{2}}\left(g\left(\ftl 1(\mathbf{X}_{j}),\ftl 2(\mathbf{X}_{j})\right)-\bE_{\mathbf{X}_{j}}\left[g\left(\ftl 1(\mathbf{X}_{j}),\ftl 2(\mathbf{X}_{j})\right)\right]\right)\right) & = & \sum_{l\in\bar{l}}w(l)o_{P}(1)\\
 & = & o_{P}(1),
\end{eqnarray*}
where $o_{P}(1)$ denotes convergence to zero in probability. In the
last step, we used the fact that if two random variables converge
in probability to constants, then their linear combination converges
in probability to the linear combination of the constants. Combining
this result with Slutsky's theorem completes the proof.

\section{Proof of Theorem~\ref{thm:uniform} (Uniform MSE)}

\label{sec:UniformProof}Since the MSE is equal to the square of the
bias plus the variance, we can upper bound the left hand side of (\ref{eq:mse_uniform})
with 
\begin{align*}
\sup_{p,q\in\Sigma(s,K_{H},\epsilon_{0},\epsilon_{\infty})}\mathbb{E}\left[\left(\tilde{\mathbf{G}}_{w_{0}}-G(p,q)\right)^{2}\right] & =\sup_{p,q\in\Sigma(s,K_{H},\epsilon_{0},\epsilon_{\infty})}\left(\text{Bias}\left(\tilde{\mathbf{G}}_{w_{0}}\right)^{2}+\text{Var}\left(\tilde{\mathbf{G}}_{w_{0}}\right)\right)\\
 & \leq\sup_{p,q\in\Sigma(s,K_{H},\epsilon_{0},\epsilon_{\infty})}\text{Bias}\left(\tilde{\mathbf{G}}_{w_{0}}\right)^{2}+\sup_{p,q\in\Sigma(s,K_{H},\epsilon_{0},\epsilon_{\infty})}\text{Var}\left(\tilde{\mathbf{G}}_{w_{0}}\right).
\end{align*}

From the assumptions (lipschitz, kernel bounded, weight calculated
from relaxed opt. prob), we have that 
\begin{eqnarray*}
\sup_{p,q\in\Sigma(s,K_{H},\epsilon_{0},\epsilon_{\infty})}\text{Var}\left(\tilde{\mathbf{G}}_{w_{0}}\right) & \leq & \sup_{p,q\in\Sigma(s,K,\epsilon_{0},\epsilon_{\infty})}\frac{11C_{g}^{2}||w_{0}||_{2}^{2}||K||_{\infty}}{N}\\
 & = & \frac{11C_{g}^{2}||w_{0}||_{2}^{2}||K||_{\infty}}{N},
\end{eqnarray*}
 where the last step follows from the fact that all of the terms are
independent of $p$ and $q$.

For the bias, recall that if $g$ is infinitely differentiable and
if the optimal weight $w_{0}$ is calculated using the relaxed convex
optimization problem, then 
\begin{eqnarray}
\text{Bias}\left(\tilde{\mathbf{G}}_{w_{0}}\right) & = & \sum_{i\in J}c_{i}(p,q)\epsilon N^{-1/2},\nonumber \\
\implies\text{Bias}\left(\tilde{\mathbf{G}}_{w_{0}}\right)^{2} & = & \frac{\epsilon^{2}}{N}\left(\sum_{i\in J}c_{i}(p,q)\right)^{2}.\label{eq:bias_sq}
\end{eqnarray}
 We use a topology argument to bound the supremum of this term. We
will use the Extreme Value Theorem~\cite{munkres2000topology}: 

\begin{theorem}[Extreme Value Theorem]\emph{Let $f:X\rightarrow\mathbb{R}$
be continuous. If $X$ is compact, then for every $x\in X$, there
exist points $c,d\in X$ s.t. $f(c)\leq f(x)\leq f(d)$.} \end{theorem} 

By this theorem, $f$ achieves its minimum and maximum on $X$. Our
approach is to first show that the functionals $c_{i}(p,q)$ are continuous
wrt $p$ and $q$ in some appropriate norm. We will then show that
the space $\Sigma(s,K_{H},\epsilon_{0},\epsilon_{\infty})$ is compact
wrt this norm. The Extreme Value Theorem can then be applied to bound
the supremum of (\ref{eq:bias_sq}).

We first define the norm. Let $\alpha=s-r>0$. We use the standard
norm on the space $\Sigma(s,K_{H})$ \cite{evans2010partial}: 
\begin{eqnarray*}
||f|| & = & ||f||_{\Sigma(s,K_{H})}\\
 & = & ||f||_{C^{r}}+\max_{|\beta|=r}|D^{\beta}f|_{C^{0,\alpha}}
\end{eqnarray*}
 where 
\begin{eqnarray*}
||f||_{C^{r}} & = & \max_{|\beta|\leq r}\sup_{x\in\mathcal{S}}|D^{\beta}f(x)|,\\
|f|_{C^{0,\alpha}} & = & \sup_{x\neq y\in\mathcal{S}}\frac{|f(x)-f(y)|}{|x-y|^{\alpha}}.
\end{eqnarray*}

\begin{lemma} \emph{The functionals $c_{m}(p,q)$ are continuous
wrt the norm $\max(||p||_{C^{r}},||q||_{C^{r}})$.} \end{lemma} 
\begin{IEEEproof}
The functionals $c_{m}(p,q)$ depend on terms of the form 
\begin{equation}
c(p,q)=\int\left(\left.\frac{\partial^{i+j}g(t_{1},t_{2})}{\partial t_{1}^{i}\partial t_{2}^{j}}\right|_{\begin{array}{c}
t_{1}=p(x)\\
t_{2}=q(x)
\end{array}}\right)D^{\beta}p(x)D^{\gamma}q(x)q(x)dx.\label{eq:const_cont}
\end{equation}
 It is sufficient to show that this is continuous. Let $\epsilon>0$
and $\max\left(||p-p_{0}||_{C^{r}},||q-q_{0}||_{C^{r}}\right)<\delta$
where $\delta>0$ will be chosen later. Then by applying the triangle
inequality for integration and adding and subtracting terms, we have
that 
\[
|c(p,q)-c(p_{0},q_{0})|
\]
 
\begin{align}
 & \leq & \int\left|\left(\left.\frac{\partial^{i+j}g(t_{1},t_{2})}{\partial t_{1}^{i}\partial t_{2}^{j}}\right|_{\begin{array}{c}
t_{1}=p(x)\\
t_{2}=q(x)
\end{array}}\right)D^{\beta}p(x)D^{\gamma}q(x)\left(q(x)-q_{0}(x)\right)\right|dx\nonumber \\
 &  & +\int\left|\left(\left.\frac{\partial^{i+j}g(t_{1},t_{2})}{\partial t_{1}^{i}\partial t_{2}^{j}}\right|_{\begin{array}{c}
t_{1}=p(x)\\
t_{2}=q(x)
\end{array}}\right)D^{\beta}p(x)q_{0}(x)\left(D^{\gamma}q(x)-D^{\gamma}q_{0}(x)\right)\right|dx\nonumber \\
 &  & +\int\left|D^{\beta}p_{0}(x)D^{\gamma}q_{0}(x)q_{0}(x)\left(\left(\left.\frac{\partial^{i+j}g(t_{1},t_{2})}{\partial t_{1}^{i}\partial t_{2}^{j}}\right|_{\begin{array}{c}
t_{1}=p(x)\\
t_{2}=q(x)
\end{array}}\right)\right.\right.\nonumber \\
 &  & -\left.\left.\left(\left.\frac{\partial^{i+j}g(t_{1},t_{2})}{\partial t_{1}^{i}\partial t_{2}^{j}}\right|_{\begin{array}{c}
t_{1}=p_{0}(x)\\
t_{2}=q_{0}(x)
\end{array}}\right)\right)\right|dx\nonumber \\
 &  & +\int\left|\left(\left.\frac{\partial^{i+j}g(t_{1},t_{2})}{\partial t_{1}^{i}\partial t_{2}^{j}}\right|_{\begin{array}{c}
t_{1}=p(x)\\
t_{2}=q(x)
\end{array}}\right)D^{\gamma}q_{0}(x)q_{0}(x)\left(D^{\beta}p(x)-D^{\beta}p_{0}(x)\right)\right|dx.\label{eq:addzero}
\end{align}
 By Assumption $\mathcal{A}.4$, the absolute value of the mixed derivatives
of $g$ is bounded on the range defined for $p$ and $q$ by some
constant $C_{i,j}$. Also, $q_{0}(x)\leq\epsilon_{\infty}$. Furthermore,
since $D^{\gamma}q_{0}$ and $D^{\beta}p$ are continuous, and since
$\mathcal{S}\subset\mathbb{R}^{d}$ is compact, then the absolute
value of the derivatives $D^{\gamma}q_{0}$ and $D^{\beta}p$ is also
bounded by a constant $\epsilon_{\infty}^{'}$. Let $\delta_{0}>0$.
Then since the mixed derivatives of $g$ are continuous on the interval
$[\epsilon_{0},\epsilon_{\infty}]$, they are uniformly continuous.
Therefore, we can choose $\delta$ small enough s.t.
\begin{equation}
\left|\left(\left.\frac{\partial^{i+j}g(t_{1},t_{2})}{\partial t_{1}^{i}\partial t_{2}^{j}}\right|_{\begin{array}{c}
t_{1}=p(x)\\
t_{2}=q(x)
\end{array}}\right)-\left(\left.\frac{\partial^{i+j}g(t_{1},t_{2})}{\partial t_{1}^{i}\partial t_{2}^{j}}\right|_{\begin{array}{c}
t_{1}=p_{0}(x)\\
t_{2}=q_{0}(x)
\end{array}}\right)\right|<\delta_{0}.\label{eq:gderivatives}
\end{equation}
 Combining all of these results with (\ref{eq:addzero}) gives 
\begin{eqnarray*}
|c(p,q)-c(p_{0},q_{0})| & \leq & \lambda(\mathcal{S})\delta C_{ij}\epsilon_{\infty}^{'}\left(2+\epsilon_{\infty}\right)\\
 &  & +\lambda(\mathcal{S})\epsilon_{\infty}^{'}\epsilon_{\infty}\left(2\delta_{0}+C_{ij}\delta\right),
\end{eqnarray*}
 where $\lambda(\mathcal{S})$ is the Lebesgue measure of $\mathcal{S}$.
This is bounded since $\mathcal{S}$ is compact. Let $\delta_{0}^{'}>0$
be s.t. if $\max\left(||p-p_{0}||_{C^{r}},||q-q_{0}||_{C^{r}}\right)<\delta_{0}^{'}$,
then (\ref{eq:gderivatives}) is less than $\frac{\epsilon}{4\lambda(\mathcal{S})\epsilon_{\infty}^{'}\epsilon_{\infty}}$.
Let $\delta_{1}=\frac{\epsilon}{4\lambda(\mathcal{S})C_{ij}\epsilon_{\infty}^{'}(1+\epsilon_{\infty})}$.
Then if $\delta<\min(\delta_{0}^{'},\delta_{1})$, then 
\[
|c(p,q)-c(p_{0},q_{0})|<\epsilon.
\]
\end{IEEEproof}
Given that each $c_{i}(p,q)$ is continuous, then $\left(\sum_{i\in J}c_{i}(p,q)\right)^{2}$
is also continuous wrt $p$ and $q$.

We now argue that $\Sigma(s,K_{H})$ is compact. First, a set is relatively
compact if its closure is compact. By the Arzela-Ascoli theorem~\cite{gilbarg2001elliptic},
the space $\Sigma(s,K_{H})$ is relatively compact in the topology
induced by the $\left\Vert \cdot\right\Vert _{\Sigma(t,K_{H})}$ norm
for any $t<s$. We choose $t=r$. It can then be shown that under
the $\left\Vert \cdot\right\Vert _{\Sigma(r,K_{H})}$ norm, $\Sigma(s,K_{H})$
is complete~\cite{evans2010partial}. Since $\Sigma(s,K_{H})$ is
contained in a metric space, then it is also closed and therefore
equal to its closure. Thus $\Sigma(s,K_{H})$ is compact. Then since
$\Sigma(s,K_{H},\epsilon_{0},\epsilon_{\infty})$ is closed in $\Sigma(s,K_{H})$,
it is also compact. Therefore, since for each $p,q\in\Sigma(s,K_{H},\epsilon_{0},\epsilon_{\infty})$,
$\left(\sum_{i\in J}c_{i}(p,q)\right)^{2}<\infty$, by the Extreme
Value Theorem we have 
\begin{eqnarray*}
\sup_{p,q\in\Sigma(s,K_{H},\epsilon_{0},\epsilon_{\infty})}\text{Bias}\left(\tilde{\mathbf{G}}_{w_{0}}\right)^{2} & = & \sup_{p,q\in\Sigma(s,K_{H},\epsilon_{0},\epsilon_{\infty})}\frac{\epsilon^{2}}{N}\left(\sum_{i\in J}c_{i}(p,q)\right)^{2}\\
 & = & \frac{\epsilon^{2}}{N}C,
\end{eqnarray*}
 where we use the fact that $J$ is finite (see Sections~\ref{sub:odin}
or \ref{sec:Mod}for the set $J$ ).

\bibliographystyle{IEEEbib}
\bibliography{References}

\begin{thebibliography}{10}

\bibitem{moon2016improving}
Kevin~R Moon, Kumar Sricharan, Kristjan Greenewald, and Alfred~O Hero,
\newblock ``Improving convergence of divergence functional ensemble
  estimators,''
\newblock in {\em Information Theory (ISIT), 2016 IEEE International Symposium
  on}. IEEE, 2016, pp. 1133--1137.

\bibitem{cover2012elements}
T.~M. Cover and J.~A. Thomas,
\newblock {\em Elements of information theory},
\newblock John Wiley \& Sons, 2012.

\bibitem{avi1996bound}
H.~Avi-Itzhak and T.~Diep,
\newblock ``Arbitrarily tight upper and lower bounds on the {Bayesian}
  probability of error,''
\newblock {\em IEEE Transactions on Pattern Analysis and Machine Intelligence},
  vol. 18, no. 1, pp. 89--91, 1996.

\bibitem{hashlamoun1994bound}
W.~A. Hashlamoun, P.~K. Varshney, and V.~Samarasooriya,
\newblock ``A tight upper bound on the {Bayesian} probability of error,''
\newblock {\em IEEE Transactions on Pattern Analysis and Machine Intelligence},
  vol. 16, no. 2, pp. 220--224, 1994.

\bibitem{moon2015Bayes}
K.R. Moon, V.~Delouille, and A.~O. Hero~III,
\newblock ``Meta learning of bounds on the {Bayes} classifier error,''
\newblock in {\em IEEE Signal Processing and SP Education Workshop}. IEEE,
  2015, pp. 13--18.

\bibitem{chernoff1952measure}
H.~Chernoff,
\newblock ``A measure of asymptotic efficiency for tests of a hypothesis based
  on the sum of observations,''
\newblock {\em The Annals of Mathematical Statistics}, pp. 493--507, 1952.

\bibitem{berisha2014bound}
V.~Berisha, A.~Wisler, A.~O. Hero~III, and A.~Spanias,
\newblock ``Empirically estimable classification bounds based on a new
  divergence measure,''
\newblock {\em IEEE Transactions on Signal Processing}, 2015.

\bibitem{moon2014nips}
K.~R. Moon and A.~O. Hero~III,
\newblock ``Multivariate f-divergence estimation with confidence,''
\newblock in {\em Advances in Neural Information Processing Systems}, 2014, pp.
  2420--2428.

\bibitem{gliske2015intrinsic}
Stephen~V Gliske, Kevin~R Moon, William~C Stacey, and Alfred~O Hero~III,
\newblock ``The intrinsic value of {HFO} features as a biomarker of epileptic
  activity,''
\newblock in {\em IEEE International Conference on Acoustics, Speech, and
  Signal Processing}, 2016.

\bibitem{poczos2011estimation}
B.~P{\'o}czos and J.~G. Schneider,
\newblock ``On the estimation of alpha-divergences,''
\newblock in {\em International Conference on Artificial Intelligence and
  Statistics}, 2011, pp. 609--617.

\bibitem{oliva2013distribution}
J.~Oliva, B.~P{\'o}czos, and J.~Schneider,
\newblock ``Distribution to distribution regression,''
\newblock in {\em Proceedings of The 30th International Conference on Machine
  Learning}, 2013, pp. 1049--1057.

\bibitem{szabo2014distribution}
Z.~Szab{\'o}, A.~Gretton, B.~P{\'o}czos, and B.~Sriperumbudur,
\newblock ``Two-stage sampled learning theory on distributions,''
\newblock {\em To appear in AISTATS}, 2015.

\bibitem{moon2015partII}
K.~R. Moon, V.~Delouille, J.~J. Li, R.~De~Visscher, F.~Watson, and A.~O.
  Hero~III,
\newblock ``Image patch analysis of sunspots and active regions. {II.
  Clustering} via matrix factorization,''
\newblock {\em Journal of Space Weather and Space Climate}, vol. 6, no. A3,
  2016.

\bibitem{moon2015partI}
K.~R. Moon, J.~J. Li, V.~Delouille, R.~De~Visscher, F.~Watson, and A.~O.
  Hero~III,
\newblock ``Image patch analysis of sunspots and active regions. {I. Intrinsic}
  dimension and correlation analysis,''
\newblock {\em Journal of Space Weather and Space Climate}, vol. 6, no. A2,
  2016.

\bibitem{dhillon2003cluster}
I.~S. Dhillon, S.~Mallela, and R.~Kumar,
\newblock ``A divisive information theoretic feature clustering algorithm for
  text classification,''
\newblock {\em The Journal of Machine Learning Research}, vol. 3, pp.
  1265--1287, 2003.

\bibitem{banerjee2005clustering}
A.~Banerjee, S.~Merugu, I.~S. Dhillon, and J.~Ghosh,
\newblock ``Clustering with {Bregman} divergences,''
\newblock {\em The Journal of Machine Learning Research}, vol. 6, pp.
  1705--1749, 2005.

\bibitem{lewi2006real}
J.~Lewi, R.~Butera, and L.~Paninski,
\newblock ``Real-time adaptive information-theoretic optimization of
  neurophysiology experiments,''
\newblock in {\em Advances in Neural Information Processing Systems}, 2006, pp.
  857--864.

\bibitem{bruzzone1995feature}
L.~Bruzzone, F.~Roli, and S.~B. Serpico,
\newblock ``An extension of the {Jeffreys-Matusita} distance to multiclass
  cases for feature selection,''
\newblock {\em Geoscience and Remote Sensing, IEEE Transactions on}, vol. 33,
  no. 6, pp. 1318--1321, 1995.

\bibitem{guorong1996feature}
X.~Guorong, C.~Peiqi, and W.~Minhui,
\newblock ``Bhattacharyya distance feature selection,''
\newblock in {\em Pattern Recognition, 1996., Proceedings of the 13th
  International Conference on}. IEEE, 1996, vol.~2, pp. 195--199.

\bibitem{sakate2014variable}
D.~M. Sakate and D.~N. Kashid,
\newblock ``Variable selection via penalized minimum $\varphi$-divergence
  estimation in logistic regression,''
\newblock {\em Journal of Applied Statistics}, vol. 41, no. 6, pp. 1233--1246,
  2014.

\bibitem{hild2001blind}
K.~E. Hild, D.~Erdogmus, and J.~C. Principe,
\newblock ``Blind source separation using {Renyi's} mutual information,''
\newblock {\em Signal Processing Letters, IEEE}, vol. 8, no. 6, pp. 174--176,
  2001.

\bibitem{mihoko2002blind}
M.~Mihoko and S.~Eguchi,
\newblock ``Robust blind source separation by beta divergence,''
\newblock {\em Neural computation}, vol. 14, no. 8, pp. 1859--1886, 2002.

\bibitem{vemuri2011segment}
B.~C. Vemuri, M.~Liu, S.~Amari, and F.~Nielsen,
\newblock ``Total {Bregman divergence and its applications to DTI} analysis,''
\newblock {\em Medical Imaging, IEEE Transactions on}, vol. 30, no. 2, pp.
  475--483, 2011.

\bibitem{hamza2003segmentation}
A.~B. Hamza and H.~Krim,
\newblock ``Image registration and segmentation by maximizing the
  {Jensen-R}{\'e}nyi divergence,''
\newblock in {\em Energy Minimization Methods in Computer Vision and Pattern
  Recognition}. Springer, 2003, pp. 147--163.

\bibitem{liu2014segment}
G.~Liu, G.~Xia, W.~Yang, and N.~Xue,
\newblock ``{SAR} image segmentation via non-local active contours,''
\newblock in {\em Geoscience and Remote Sensing Symposium (IGARSS), 2014 IEEE
  International}. IEEE, 2014, pp. 3730--3733.

\bibitem{korzhik2015steganographic}
V.~Korzhik and I.~Fedyanin,
\newblock ``Steganographic applications of the nearest-neighbor approach to
  {Kullback-Leibler} divergence estimation,''
\newblock in {\em Digital Information, Networking, and Wireless Communications
  (DINWC), 2015 Third International Conference on}. IEEE, 2015, pp. 133--138.

\bibitem{basseville2013divergence}
M.~Basseville,
\newblock ``Divergence measures for statistical data processing--{An} annotated
  bibliography,''
\newblock {\em Signal Processing}, vol. 93, no. 4, pp. 621--633, 2013.

\bibitem{csiszar1967div}
I.~Csiszar,
\newblock ``Information-type measures of difference of probability
  distributions and indirect observations,''
\newblock {\em Studia Sci. MAth. Hungar.}, vol. 2, pp. 299--318, 1967.

\bibitem{ali1966div}
S.~M. Ali and S.~D. Silvey,
\newblock ``A general class of coefficients of divergence of one distribution
  from another,''
\newblock {\em Journal of the Royal Statistical Society. Series B
  (Methodological)}, pp. 131--142, 1966.

\bibitem{kullback1951divergence}
S.~Kullback and R.~A. Leibler,
\newblock ``On information and sufficiency,''
\newblock {\em The Annals of Mathematical Statistics}, vol. 22, no. 1, pp.
  79--86, 1951.

\bibitem{renyi1961divergence}
A.~R{\'e}nyi,
\newblock ``On measures of entropy and information,''
\newblock in {\em Fourth Berkeley Sympos. on Mathematical Statistics and
  Probability}, 1961, pp. 547--561.

\bibitem{hellinger1909}
E.~Hellinger,
\newblock ``{Neue Begr{\"u}ndung der Theorie quadratischer Formen von
  unendlichvielen Ver}{\"a}nderlichen.,''
\newblock {\em Journal f{\"u}r die reine und angewandte Mathematik}, vol. 136,
  pp. 210--271, 1909.

\bibitem{bhattacharyya1946div}
A.~Bhattacharyya,
\newblock ``On a measure of divergence between two multinomial populations,''
\newblock {\em Sankhy{\=a}: The Indian Journal of Statistics}, pp. 401--406,
  1946.

\bibitem{wang2009divergence}
Qing Wang, Sanjeev~R Kulkarni, and Sergio Verd{\'u},
\newblock ``Divergence estimation for multidimensional densities via
  k-nearest-neighbor distances,''
\newblock {\em IEEE Trans. Information Theory}, vol. 55, no. 5, pp. 2392--2405,
  2009.

\bibitem{darbellay1999MIest}
Georges~A Darbellay, Igor Vajda, et~al.,
\newblock ``Estimation of the information by an adaptive partitioning of the
  observation space,''
\newblock {\em IEEE Trans. Information Theory}, vol. 45, no. 4, pp. 1315--1321,
  1999.

\bibitem{silva2010partition}
Jorge Silva and Shrikanth~S Narayanan,
\newblock ``Information divergence estimation based on data-dependent
  partitions,''
\newblock {\em Journal of Statistical Planning and Inference}, vol. 140, no.
  11, pp. 3180--3198, 2010.

\bibitem{le2013partition}
Trung~Kien Le,
\newblock ``Information dependency: {Strong consistency of Darbellay--Vajda}
  partition estimators,''
\newblock {\em Journal of Statistical Planning and Inference}, vol. 143, no.
  12, pp. 2089--2100, 2013.

\bibitem{wang2005part}
Qing Wang, Sanjeev~R Kulkarni, and Sergio Verd{\'u},
\newblock ``Divergence estimation of continuous distributions based on
  data-dependent partitions,''
\newblock {\em IEEE Trans. Information Theory}, vol. 51, no. 9, pp. 3064--3074,
  2005.

\bibitem{hero2002applications}
A.~O. Hero~III, B.~Ma, O.~Michel, and J.~Gorman,
\newblock ``Applications of entropic spanning graphs,''
\newblock {\em Signal Processing Magazine, IEEE}, vol. 19, no. 5, pp. 85--95,
  2002.

\bibitem{moon2014isit}
K.~R. Moon and A.~O. Hero~III,
\newblock ``Ensemble estimation of multivariate f-divergence,''
\newblock in {\em Information Theory (ISIT), 2014 IEEE International Symposium
  on}. IEEE, 2014, pp. 356--360.

\bibitem{nguyen2010div}
X.~Nguyen, M.~J. Wainwright, and M.~I. Jordan,
\newblock ``Estimating divergence functionals and the likelihood ratio by
  convex risk minimization,''
\newblock {\em Information Theory, IEEE Transactions on}, vol. 56, no. 11, pp.
  5847--5861, 2010.

\bibitem{krishnamurthy2014divergence}
A.~Krishnamurthy, K.~Kandasamy, B.~Poczos, and L.~Wasserman,
\newblock ``Nonparametric estimation of renyi divergence and friends,''
\newblock in {\em Proceedings of The 31st International Conference on Machine
  Learning}, 2014, pp. 919--927.

\bibitem{singh2014renyi}
S.~Singh and B.~P{\'o}czos,
\newblock ``Generalized exponential concentration inequality for r{\'e}nyi
  divergence estimation,''
\newblock in {\em Proceedings of the 31st International Conference on Machine
  Learning (ICML-14)}, 2014, pp. 333--341.

\bibitem{singh2014exponential}
S.~Singh and B.~P{\'o}czos,
\newblock ``Exponential concentration of a density functional estimator,''
\newblock in {\em Advances in Neural Information Processing Systems}, 2014, pp.
  3032--3040.

\bibitem{kandasamy2015nonparametric}
Kirthevasan Kandasamy, Akshay Krishnamurthy, Barnabas Poczos, Larry Wasserman,
  and James Robins,
\newblock ``Nonparametric von mises estimators for entropies, divergences and
  mutual informations,''
\newblock in {\em Advances in Neural Information Processing Systems}, 2015, pp.
  397--405.

\bibitem{hardle1990applied}
Wolfgang H{\"a}rdle,
\newblock {\em Applied Nonparametric Regression},
\newblock Cambridge University Press, 1990.

\bibitem{berlinet1995l1}
A~Berlinet, L~Devroye, and L~Gy{\"o}rfi,
\newblock ``Asymptotic normality of {L1} error in density estimation,''
\newblock {\em Statistics}, vol. 26, pp. 329--343, 1995.

\bibitem{berlinet1997asymptotic}
A~Berlinet, L{\'a}szl{\'o} Gy{\"o}rfi, and Istv{\'a}n D{\'e}nes,
\newblock ``Asymptotic normality of relative entropy in multivariate density
  estimation,''
\newblock {\em Publications de l'Institut de Statistique de l'Universit{\'e} de
  Paris}, vol. 41, pp. 3--27, 1997.

\bibitem{bickel1973some}
Peter~J Bickel and Murray Rosenblatt,
\newblock ``On some global measures of the deviations of density function
  estimates,''
\newblock {\em The Annals of Statistics}, pp. 1071--1095, 1973.

\bibitem{sricharan2013ensemble}
K.~Sricharan, D.~Wei, and A.~O. Hero,
\newblock ``Ensemble estimators for multivariate entropy estimation,''
\newblock {\em Information Theory, IEEE Transactions on}, vol. 59, no. 7, pp.
  4374--4388, 2013.

\bibitem{hansen2009lecture}
Bruce~E Hansen,
\newblock ``Lecture notes on nonparametrics,''
\newblock 2009.

\bibitem{efron1981jackknife}
Bradley Efron and Charles Stein,
\newblock ``The jackknife estimate of variance,''
\newblock {\em The Annals of Statistics}, pp. 586--596, 1981.

\bibitem{moon2016arxiv}
Kevin~R Moon, Kumar Sricharan, Kristjan Greenewald, and Alfred~O Hero~III,
\newblock ``Nonparametric ensemble estimation of distributional functionals,''
\newblock {\em arXiv preprint arXiv:1601.06884v2}, 2016.

\bibitem{paul2015transcriptional}
Franziska Paul, Ya'ara Arkin, Amir Giladi, Diego~Adhemar Jaitin, Ephraim
  Kenigsberg, Hadas Keren-Shaul, Deborah Winter, David Lara-Astiaso, Meital
  Gury, Assaf Weiner, et~al.,
\newblock ``Transcriptional heterogeneity and lineage commitment in myeloid
  progenitors,''
\newblock {\em Cell}, vol. 163, no. 7, pp. 1663--1677, 2015.

\bibitem{kanehisa2000kegg}
Minoru Kanehisa and Susumu Goto,
\newblock ``Kegg: kyoto encyclopedia of genes and genomes,''
\newblock {\em Nucleic acids research}, vol. 28, no. 1, pp. 27--30, 2000.

\bibitem{kanehisa2015kegg}
Minoru Kanehisa, Yoko Sato, Masayuki Kawashima, Miho Furumichi, and Mao Tanabe,
\newblock ``Kegg as a reference resource for gene and protein annotation,''
\newblock {\em Nucleic acids research}, vol. 44, no. D1, pp. D457--D462, 2015.

\bibitem{kanehisa2016kegg}
Minoru Kanehisa, Miho Furumichi, Mao Tanabe, Yoko Sato, and Kanae Morishima,
\newblock ``Kegg: new perspectives on genomes, pathways, diseases and drugs,''
\newblock {\em Nucleic acids research}, vol. 45, no. D1, pp. D353--D361, 2016.

\bibitem{van2017magic}
David van Dijk, Juozas Nainys, Roshan Sharma, Pooja Kathail, Ambrose~J Carr,
  Kevin~R Moon, Linas Mazutis, Guy Wolf, Smita Krishnaswamy, and Dana Pe'er,
\newblock ``Magic: A diffusion-based imputation method reveals gene-gene
  interactions in single-cell rna-sequencing data,''
\newblock {\em bioRxiv}, p. 111591, 2017.

\bibitem{durrett2010probability}
Rick Durrett,
\newblock {\em Probability: Theory and Examples},
\newblock Cambridge University Press, 2010.

\bibitem{gut2012probability}
Allan Gut,
\newblock {\em Probability: A Graduate Course},
\newblock Springer Science \& Business Media, 2012.

\bibitem{munkres2000topology}
James Munkres,
\newblock {\em Topology},
\newblock Prentice Hall, Inc, 2000.

\bibitem{evans2010partial}
Lawrence~C Evans,
\newblock {\em Partial differential equations},
\newblock American Mathematical Society, 2010.

\bibitem{gilbarg2001elliptic}
David Gilbarg and Neil~S Trudinger,
\newblock {\em Elliptic partial differential equations of second order},
\newblock Springer, 2001.

\end{thebibliography}

\end{document}